\documentclass[fleqn,usenatbib]{mnras}

\usepackage{newtxtext,newtxmath}

\usepackage[T1]{fontenc}

\DeclareRobustCommand{\VAN}[3]{#2}
\let\VANthebibliography\thebibliography
\def\thebibliography{\DeclareRobustCommand{\VAN}[3]{##3}\VANthebibliography}

\usepackage{graphicx}	

\usepackage{txfonts}
\usepackage{mathtools}
\usepackage[]{natbib}

\usepackage{rotating}
\usepackage{amssymb}
\usepackage{longtable}
\usepackage{epsfig}
\usepackage{txfonts} 
\usepackage{natbib} 
\bibpunct{(}{)}{;}{a}{}{,} 
\usepackage{lscape}
\usepackage{times}
\usepackage{txfonts}
\usepackage[T1]{fontenc}

\usepackage[font=small,labelfont=bf,tableposition=top]{caption} 
\usepackage{booktabs} 
\usepackage{float}
\usepackage{textcomp}
\usepackage{euscript}
\usepackage{multicol} 
\usepackage{caption} 
\usepackage{blindtext}
\usepackage{lipsum} 
\usepackage{stfloats}

\usepackage[T1]{fontenc}
\usepackage[utf8]{inputenc}

\usepackage[figurename=Figure.,
                  justification=RaggedRight, 
                  labelfont={bf, footnotesize}, 
                  textfont={footnotesize},position=top]{caption}

\usepackage{hyperref}
\title[Gas-grain cycling and the synthesis of NH$_3$]{Species cycling and the enhancement of ammonia in prestellar cores}

\author[A. A. von Prochazka et al.]{
Azrael A. von Proch\'{a}zka\href{https://orcid.org/0000-0001-9286-9026}{\includegraphics[scale=0.065]{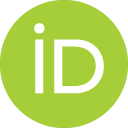}}$^{1}$\thanks{E-mail: avonprochazka01@qub.ac.uk},
T. J. Millar\href{https://orcid.org/0000-0001-5178-3656}{\includegraphics[scale=0.065]{ORCID-iD_icon-128x128.png}}$^{1}$
\\
$^{1}$Astrophysics Research Centre, School of Mathematics and Physics, Queen's University Belfast, Belfast, BT7 1NN, UK\\
}

\date{Accepted XXX. Received YYY; in original form ZZZ}

\pubyear{2020}

\begin{document}
\label{firstpage}
\pagerange{\pageref{firstpage}--\pageref{lastpage}}
\maketitle

\begin{abstract}
The quantity of NH$_3$ produced on grain surfaces in the prestellar core is thought to be one of the determining factors regarding the chemical complexity achievable at later stages of stellar birth. In order to investigate how this quantity might be influenced by the gas-grain cycling of molecular material within the cloud, we employ a modified rates gas-grain chemical code and follow the time-dependent chemistry of NH$_3$ as the system evolves. Our models incorporate an updated version of the most recent UDfA network of reaction rate coefficients, desorption from the grains through standard thermal and non-thermal processes, and physisorbed and chemisorbed binding of atomic and molecular hydrogen to a population of carbonaceous and siliceous grains. We find that 1.) observable abundances of NH$_3$ can exist in the gas phase of our models at early times when the N atom is derived from CN via an efficient early-time hydrocarbon chemistry, 2.) a time-dependent gradient exists in the observational agreement between different species classes in our models, consistent with possible physical substructures within the TMC-1 Cyanopolyyne Peak, and 3.) the gaseous and solid-state abundances of NH$_3$ are sensitive to the presence of gas-grain cycling within the system. Our results suggest that the degree of chemical complexity achievable at later stages of the cloud's chemical evolution is indeed influenced by the manner in which the gas-grain cycling occurs.
\end{abstract}





\section{Introduction}
NH$_3$ is potentially one of the species that determines the complexity of the organic chemistry that may be achieved in the molecular clouds associated with star formation. In particular, \citet{rodgers2001chemical} found that the degree of N/O differentiation across massive protostellar cores was largely determined by the amount of NH$_3$ in the precursor ices. They found that if the abundance of solid-state NH$_{3}$ in the prestellar ice mantles is high, then the production of complex O-bearing organic molecules in the later hot core and hot corino environments would be reduced as a result of the suppression of alkyl cation transfer reactions once the NH$_3$ ice had desorbed into the gas phase. This is due fundamentally to the large proton affinity of NH$_3$, one of the largest of interstellar molecules. When the abundance of NH$_3$ is high, NH$_3$ takes protons from  CH$_3$OH$_2^{\text{+}}$ "before"  CH$_3$OH$_2^{\text{+}}$ has had the chance to react with less abundant species in order to build larger complex organic structures. By this scheme, regions of molecular clouds with higher abundances of NH$_3$ would naturally be expected to exhibit lower abundances of complex organic molecules. Conversely, \citet{rodgers2001chemical} also found that if the abundance of NH$_{3}$ ice is low, then methyl cation transfer reactions involving CH$_3$OH$_2^{\text{+}}$ may proceed, thereby enhancing the production of larger O-bearing organic molecules such as acetic acid, acetone, methyl formate, and dimethyl ether. Since chemical differentiations across spatial regions are observed in both high-mass and low-mass star-forming environments alike \citep[e.g.,][]{little1979relative, caselli1993chemdiff, pratap1997study, remijan2004survey, spezzano2017observed}, it is intuitive to surmise that NH$_3$ likely also influences the production of more complex prebiotic molecules in lower-mass hot corinos by the same scheme.

While the molecular origins for this variance in N-rich and O-rich chemistries were initially assumed to perhaps be the result of differences in how CO and N$_2$ behaved in the solid state, \citet{bisschop2006desorption} found that the respective desorption temperatures and sticking probabilities between the two molecules are in fact quite similar, thereby suggesting in agreement with \citet{jorgensen2004molecular} that the anti-correlation of CO and N$_2$H$^{\text{+}}$ is in fact due to the freeze-out of CO. The late-time enhancement of NH$_3$, on the other hand, has been posited to perhaps be the result of a longer time-scale required for its synthesis from N$_2$ and/or late-time dynamic cycling of molecular material within the cloud, thereby leading to its frequent interpretation as a "late-time" molecule \citep[e.g.,][]{suzuki1992survey, bergin1997chemical, bergin2007cold, hirota2009search, agundez2013chemistry}. Indeed, because of their ubiquity and tendency to remain in the gas phase throughout the freeze-out event, both NH$_3$ and N$_2$H$^{\text{+}}$ are widely utilised as observational tracers of the physical and ionisation conditions of evolved prestellar cloud cores \citep{ho1983interstellar, walmsley1983ammonia, suzuki1992survey, Bergin_2002, caselli2002dense, tafalla2004internal, tafalla2006internal}. While \citet{von2013prestellar} found that observable abundances of gas-phase NH$_3$ can also be achieved at early times in gas-grain astrochemical simulations of low-mass star-forming environments when employing the RATE06 \citep{woodall2007umist} iteration of the UDfA (UMIST) database under C-rich conditions, \citet{agundez2013chemistry} also found this result when using the more recent RATE12\footnote{\href{http://udfa.net}{http://udfa.net}} release in their gas-phase models. \citet{maffucci2018astrochemical} explored the abundance of NH$_3$ under varying physical and chemical conditions with reactions from the KIDA\footnote{\href{http://kida.obs.u-bordeaux1.fr}{http://kida.obs.u-bordeaux1.fr}} database and found that the agreement of the molecule with observations could be maintained for "long times"\footnote{2 $\times$ 10$^{4-6}$ years.} independent of the values chosen for the C/O ratio, density, and rate of cosmic-ray ionisation. In agreement with the above authors, we find in the models herein that despite the traditional interpretation and use of NH$_3$ abundances as tracers of an evolved chemistry within the core, the gas-phase abundance of NH$_3$ can indeed reach reasonable agreement with observations at early times as well, especially when the molecule's production is driven by routes involving the early-time C-chains and cyanides. While many authors have investigated the effects of non-thermal desorption on the theoretical abundances of species observed in prestellar cores, \citep[e.g.,][]{leger1985desorption,hasegawa1993new, willacy1994desorption, garrod2007non}, questions still remain regarding the specific pathways which lead to the observed values. Our analysis of the chemistry at late times suggests that the abundance of NH$_3$ is highly sensitive to the presence and mechanics of the gas-grain cycling of molecular material within the system.

In addition to the gas-phase chemistry, the amount of NH$_3$ in the ice may also be particularly relevant towards the production of biological precursors in the solid state. For example, \citet{caro2003uv} identified the spectroscopic IR signatures of amides, esters, and the ammonium salts of carboxylic acids in the residue of their UV-irradiated ices. More recently, \citet{bera2017mechanisms} showed that the synthesis of the nucleobases adenine and guanine can occur efficiently via UV-irradiation of ice mixtures consisting of H$_2$O, NH$_3$, and purine. They posit that due to the ion-radical nature of the mechanism, the reactions are barrierless and could reasonably proceed in low-temperature environments such as meteorites. It is the goal of astrochemical modelling to unravel the underlying chemistry of such environments so that we may not only properly interpret current observations towards these regions but also make reliable predictions and assumptions about their overarching chemistries and molecular compositions in the gas, on the grains, and in the transient state as they cycle between the two. Understanding this chemistry will hopefully allow us to answer fundamental questions regarding the formation of life in the Universe.

The physical and chemical parameters which define our models are based on those observed towards the Taurus Molecular Cloud-1 Cyanopolyyne Peak -- TMC-1 (CP), which we briefly review in Section \ref{sec:TMC-1}. We describe the parameters themselves in Section \ref{sec:Dark_Cloud_Models}. Our calculations differ from those of traditional dark cloud models in that they specifically consider: (i) the formation of H$_2$ as it might proceed across a population of realistic carbonaceous and siliceous grain types and (ii) the temperature-dependent probability that the H atom will adhere to the grains upon collision and thereafter bind through either physisorption or chemisorption. We describe the basic details for the formation of H$_2$ by this scheme in Section \ref{sec:calculations}. In Section \ref{sec:Results}, we present our results for the overall agreement between our modelled abundances and observations towards TMC-1 (CP) and discuss the key species and reactions involved in our modelled chemistry of NH$_3$. Finally, we review and summarise our conclusions in Section \ref{sec:Discussion}.

\section{TMC-1}
\label{sec:TMC-1}

Located in the Heiles Cloud 2 complex, TMC-1 is a widely-surveyed and and widely-studied molecular cloud \citep[e.g.,][]{elias1978study, churchwell1978molecular, suzuki1992survey, ohishi1998chemical, peng1998low, fosse2001molecular, kaifu20048, navarro2020gas}. While six young stellar object (YSO) candidates and an IRAS source (IRAS 04381+2540) are known to exist in the regions surrounding the TMC-1 filament \citep{hirahara1992mapping, gomez1993spatial}, \emph{Spitzer} observations indicate that the TMC-1 cloud is itself devoid of YSOs \citep[e.g.,][]{choi2017dynamical}. Consequentially, the majority of the TMC-1 complex is suspected to have evolved in relative isolation from the disruptive physical and radiative effects which generally accompany stellar ignition. At a distance of $\sim$140 pc, it is the nearest stellar nursery. For these reasons, TMC-1 is widely considered to be an ideal source through which we can observe and examine the fundamental physics and chemistry of the earliest stages of star formation.

%

The geometry of TMC-1 can be described to first order as a narrow (0.25 pc x 0.5 pc) filamentary ridge of quiescent gas-grain molecular material oriented along the SE-NW axis, with density increasing from the SE to the NW regions by perhaps a factor of 2 to as much as an order of magnitude \citep{hirahara1992mapping, pratap1997study, olano1988relative}. Common densities reported for the different regions are on the order of 10$^{4}$ cm$^{-3}$ in the SE and 10$^{5}$ cm$^{-3}$ in the NW, and the density gradient itself is often interpreted as an indication that the material in the NW is more physically evolved than that in the SE. The observations of \citet{langer1995study} and \citet{hirahara1992mapping} showed that the TMC-1 filament is in fact highly fragmented and can be distinguished into at least 6 cores with half-power radii ranging from $\sim$0.02 -- 0.05 pc. Due to the narrow line widths of the detected CCS profiles, the results of these earlier surveys also suggested that the effects of turbulence and rotation within the cloud were insignificant. However, a higher-sensitivity survey performed by \citet{peng1998low} further resolved the Cyanopolyyne Peak \citep["Core D" in][]{langer1995study, hirahara1992mapping} into at least 45 smaller cores -- of which 19 were determined to be gravitationally unbound, 21 to be stable, and 5 to be likely on the verge of self-gravitational collapse. Due to the prevalence of the unbound cores, \citet{peng1998low} suggest that these objects are not entirely transient in nature and propose microturbulence as a mechanism by which the cores are able to remain pressure-bound within the interclump gas. More recently, \citet{dobashi2018spectral, dobashi2019discovery} identified 21 different velocity-coherent substructures within TMC-1 and found that not only is their typical detected CCS line profile in the Cyanopolyyne Peak best fit by four different Gaussian components (thereby implying the presence of substructures travelling at four different velocities), but the TMC-1 complex is itself indeed likely to be in the early stages of self-gravitational collapse.

In addition to the density gradient, there also exists a chemical gradient along the SE-NW axis of TMC-1 which is commonly referred to as the "chemical ridge" \citep{olano1988relative, hirahara1992mapping, suzuki1992survey, pratap1997study, hirota2009search}. The lower-density Cyanopolyyne Peak in the SE region demonstrates a general prominence of complex carbon chain species -- notably cyanopolyynes such as HC$_3$N -- whereas the higher-density material in the NW (the "NH$_3$ Peak") demonstrates higher column densities of NH$_3$, N$_2$H$^{\text{+}}$, and SO. More recent observations have shown that a gradient of the CH fractional abundance exists along the TMC-1 ridge as well \citep{suutarinen2011ch}. While \citet{hirahara1992mapping}, \citet{suzuki1992survey}, and \citet{van1993chemical} proposed that this chemical gradient is perhaps the result of a gradient in the duration of chemical evolution or a gradient in the physical conditions along the filament, \citet{pratap1997study} suggested that the gradient could also be caused by differences in the C/O ratio between the two peaks. According to the theory, a younger region of molecular material such as the Cyanopolyyne Peak would naturally be expected to have more ambient C with which to drive an enhanced production of assorted C-bearing species. In older regions, on the other hand, most of the carbon would be expected to already be "locked" within the resilient CO molecule, whereas the NH$_3$ abundance would be expected to be high as a result of its derivation from N$_2$, which is itself thought to be synthesised by slow neutral-neutral reactions in the gas phase. These theories are in agreement with the chemical and dynamical ages of material along the TMC-1 ridge, which are often estimated to be $\sim$2 $\times$ 10$^5$ years towards the SE and at least $\sim$10$^{6}$ to around 10$^{7}$ years towards the NW \citep[e.g.,][]{suzuki1992survey, hirahara1992mapping, pratap1997study, olano1988relative, saito2002chemical, suutarinen2011ch}. As a possible explanation for the chemical gradient across the filament, \citet{markwick2000abundance} investigated the effects of MHD waves propagating from IRAS 04381+2540, while \citet{hartquist2001chemical} considered that the observed microstructure could be caused by a stellar-wind-induced shock that was slow enough so as not to activate a high-temperature chemistry within the core. Likewise, \citet{dickens2001small} considered that the gradient could be induced by MHD waves resulting from clump-clump collisions or, alternately, direct grain-grain collisions resulting from the impact of two colliding clumps within the medium. More recently, \citet{choi2017dynamical} performed a study of the 1.2-mm continuum emission in TMC-1 which suggests that the Cyanopolyyne Peak is in fact in the process of rapid core formation due to the collision of two molecular clouds. In their analysis, they suggest that the observed chemical gradient between the different regions is likely the result of a difference in dynamical evolution, with the Cyanopolyyne Peak being formed through collision-induced free fall and the NH$_3$ Peak being formed instead through ambipolar diffusion. In this work, we focus our investigation on the chemistry of the younger, lower-density Cyanopolyyne Peak.






The gas-grain cycling of molecular material has long been anticipated to be central to the chemical modelling of dark clouds, and TMC-1 (CP) is often invoked as a reference source through which to explore potentially viable theoretical frameworks which might explain the specific nature of the underlying gas-grain chemistry. While the inclusion of species desorption from the grains through thermal evaporation ("THERM") and non-thermal processes such as cosmic-ray heating ("CRDES"), cosmic-ray-induced photodesorption ("PDD"), and the release of excess energy onto the grains by the formation of H$_2$ on the surface ("H2DES") in astrochemical simulations has been present since the models of \citet{hasegawa1992models}, \citet{willacy1993desorption}, \citet{willacy1994desorption}, and \citet{willacy1998desorption}, observations of complex organic molecules ("COMs") in the gas phase of low-temperature dense cores \citep{matthews1985acetaldehyde, friberg1988methanol, oberg2010complexchem, cernicharo2012ch3o, bacmann2012coms, vastel2014coms, soma2018coms, scibelli2020coms} requires an alternate explanation. To investigate this, \citet{garrod2007non} implemented Rice-Ramsperger-Kessel theory in order to model chemical desorption from the ices as a result of exothermic surface reactions and compared their results with observations towards TMC-1 (CP) and L134N. \citet{dulieu2013micron} observed the process experimentally in their study of water formation on silicate surfaces, and \citet{minissale2014chemdesorption} and \citet{minissale2016dust} later developed an analytical expression which describes the mechanism as it proceeds on various substrates of interstellar interest. While \citet{minissale2016dust}'s models reproduce \citet{dulieu2013micron}'s experiments well when the surfaces are bare, their results become increasingly divergent when ice mantles are present. They posit that this is the effect of a reduction in efficiency of the process when the surface coverage on the grains is high. They propose that the presence of pre-adsorbed species makes it increasingly likely that the energy released from the surface reactions will be dissipated before the product species can be desorbed into the gas phase. Citing this inefficiency, \citet{ruaud2015modelling} explored the Eley-Rideal and complex-induced reaction mechanisms as alternate routes towards the production of COMs in low-temperature dense cores. More recently, \citet{gratier2016new} released a revised compilation of reference values for observed molecular abundances towards TMC-1 (CP) based on a Bayesian statistical method to account for outlier points in the original data of \citet{ohishi1992molecular} and \citet{ohishi1998chemical}. \citet{maffucci2018astrochemical} then investigated the effects on standard astrochemical models when their derived abundances were evaluated against these new values and explored the effects on their models when Eley-Rideal and van der Waals processes were considered as well. Thus far, the traditional approach for astrochemical models has been to assume that the formation of H$_2$ occurs on a single, arbitrary grain type. In the models herein, we consider that the formation of H$_2$ occurs instead on a population of realistic grain compositions. We describe the relevant parameters for these calculations in Section \ref{sec:calculations}.

\section{Dark Cloud Models}
\label{sec:Dark_Cloud_Models}


In our models, we simulate the chemistry of a point of gas-grain molecular material in a homogeneous, isotropic dark cloud of constant temperature 10 K, density 2 $\times$ 10$^{4}$ cm$^{-3}$, visual extinction 10 mag, and cosmic-ray ionisation rate 1.3 $\times$ 10$^{-17}$ s$^{-1}$ under conditions of local thermodynamic equilibrium (LTE). These conditions correspond to the quiescent, innermost regions of typical interstellar dark cloud cores prior to stellar ignition. Over the course of the chemical evolution in our models, atoms are synthesised into molecules in both the gas phase and on grain surfaces, with the overall chemistry of the system tending to progress towards a quasi-steady state. We adopt the value of the grain radius to be 10$^{-5}$ cm (i.e., the "classical" value) and a binding site surface density on the grains of 7.9 $\times$ 10$^{14}$ sites cm$^{-2}$. These values correspond to a total of 9.9 $\times$ 10$^{5}$ binding sites per grain and a dust-to-gas number ratio of 1.33 $\times$ 10$^{-12}$. The grain types included in our calculations are amorphous carbon, silicates, and PAH particles which can possess either high ("graphitic") or low ("para-sitic")\footnote{for which the surface chemistry is dependent on the population of atoms on the grains.} barriers for the transfer of H atoms from physisorbed to chemisorbed binding sites. We assume that each of these grain types accounts for 25\% of the overall grain population and describe the overall process for the formation of H$_2$ on these different surfaces in Section \ref{sec:h2_formation}. Our binding-site parameters are based on the data of \citet{pirronello1997laboratory, pirronello1997efficiency, pirronello1999measurements}; \citet{cazaux2009hd}; and \citet{cazaux2004h2, cazaux2010erratum}. While the location of species within the ice mantle is indeed likely to be an important factor with regard to the chemistry that occurs on the grains, we do not track this parameter in our calculations. In this sense, our calculations produce "two-phase" models. 




The formulas for the different desorption mechanisms included in our models may be found in \citet{hasegawa1992models} (THERM), \citet{hasegawa1993new} (CRDES), \citet{willacy1993desorption} (PDD), and \citet{willacy1994desorption} (H2DES), and we consider the effects on the chemistry of all possible combinations of these mechanisms. For our PDD calculations, we adopt a photodesorption yield of 3 $\times$ 10$^{-3}$ \citep[as per][]{shen2004cosmic} and a photon flux of 10$^{4}$ photons cm$^{-2}$s$^{-1}$ \citep[as per][]{oberg2007photodesorption}, and we allow the efficiency parameter in the H2DES calculations, $\epsilon$, to vary between 0.001 and 0.5.\footnote{We are as yet unaware of any reported measurements of $\epsilon$.} The model upon which we focus the majority of our analysis here includes all of the above desorption mechanisms simultaneously ("ALL") and assumes a conservative H2DES efficiency of $\epsilon$ = 0.001. As our models generally maintain substantial abundances of species on the grains,\footnote{$\sim$10$^{-5}$ with respect to the the total H nucleon number density by $\sim$10$^4$ years.} we do not include the calculations for chemical desorption presented by \citet{minissale2016dust} due to their reported divergence from experimental results when applied to icy surfaces. We initialise our models with the gas-phase elemental abundances given in Table \ref{tab:InitialElementalAbunds} and calculate the abundances of atoms and molecules via the ordinary differential equations solver within ODEPACK \citep{hindmarsh1983odepack, radhakrishnan1993description}.

Our gas-phase chemical network is based on the reactions and processes included in the UDfA RATE12 database, with rate coefficient calculations as described by \citet{mcelroy2013umist}; abundances determined per the methods of \citet{pickles1977model}, \citet{hasegawa1992models}, and \citet{hasegawa1993new}; accretion calculated per \citet{rawlings1992direct}; and diffusion across (and reactions between species on) the grains as detailed in \citet{hasegawa1992models}, \citet{caselli1998proposed}, and \citet{shalabiea1998grain}, with the assumption that the energy barrier to diffusion is equivalent to half the binding energy. However, modifications to the chemical network have been made in order to reflect new measurements and calculations which have been made subsequent to the RATE12 release. Our surface chemistry network is a compilation of reactions based on those used and recommended by UDfA \citep{mcelroy2013umist}, the KIDA database \citep{wakelam2012kinetic, ruaud2015modelling}, and the OSU\footnote{\href{http://faculty.virginia.edu/ericherb/research.html}{http://faculty.virginia.edu/ericherb/research.html}} network \citep{garrod2008complex} for species which have a direct corollary in the UDfA network and are thought likely to proceed in the ISM. 

\begin{table}
\caption{Elemental fractional abundances (listed with respect to the total H nucleon number density, "H$^{-1}$") used to initialise our dark cloud models. These values are based primarily on the "Low Metal" abundances given by \citet{lee1998bistability} as well as those given by \citet{roberts2004chemistry} and \citet{woodall2007umist} and are generally accepted as "standard" conditions from which the early chemistry of interstellar dark clouds is likely to proceed. The notation x(y) represents a number of the form x $\times$ 10$^{y}$.}              
\label{tab:InitialElementalAbunds}      
\centering                                      
\begin{tabular}{llllllccc}          
\hline\hline                        
Element  & Abundance & Element & Abundance \\    
\hline                                   
H$_2$     &  4.95(-1) & O           &  1.76(-4)\\
H             & 1.00(-2) & Si$^{\text{+}}$             & 2.00(-8)\\
He           &  1.00(-1) & Fe$^{\text{+}}$  &   2.00(-8)\\
C$^{\text{+}}$     &  7.30(-5) & S$^{\text{+}}$    &  8.00(-8)\\
N             &  2.14(-5) & P$^{\text{+}}$    &  3.00(-9)\\
\hline                                             
\end{tabular}
\end{table}



%
%



\section{H$_2$ formation}
\label{sec:calculations}


While most traditional astrochemical models have treated the rate of H$_2$ formation on grain surfaces to be simply equal to half the rate at which the gas-phase H atoms collide with the surfaces, independent of grain composition \citep[e.g.,][]{roberts2000modelling, roberts2004chemistry, mcelroy2013umist}, the reality is that the efficiency of the formation of H$_2$ in the solid state is likely to be dependent on the composition of the grains on which it forms. Moreover, due to its dependency on the rate of H$_2$ formation, the rate of species desorption through the H2DES process is likely to be affected as well.




\subsection{Sticking coefficient}
\label{sec:stcoeff}

We adopt the temperature-dependent equation for the sticking coefficient of atomic H given by \citet{cazaux2009hd}, 
\begin{equation}
S_{\rm{H}} \; \text{=} \; \left[1 \; \text{+} \; 0.4 \left(\frac{T{_{\rm{g}}} \; \text{+} \; T{_{\rm{d}}}}{100}\right)^{0.5} \text{+} \; 0.2 \left(\frac{T_{\rm{g}}}{100} \right) \; \text{+} \; 0.08\left(\frac{T_{\rm{g}}}{100}\right)^{2}\right]^{-1},
\label{eqn:stcoeff}
\end{equation}
where \textit{T}$_{\rm{g}}$ and \textit{T}$_{\rm{d}}$ represent the gas and dust temperatures, respectively. Assuming that the region exists in LTE at 10 K, we derive the sticking coefficient for atomic H to be 0.83.

\subsection{H$_2$ formation on grain surfaces}
\label{sec:h2_formation}

To calculate the rate of solid-state H$_2$ formation in our models, we consider the varying surface areas and efficiencies of H$_2$ formation across the population of different carbonaceous and siliceous grain types. We calculate the probability, $P_{\rm{H}}$, that an incoming H atom from the gas phase will bind to a chemisorbed site on first contact with the grain via the equation given by \citet{cazaux2009hd},
\begin{equation}
P_{\rm{H}} \; \text{=} \; 4 \left(1 \text{+}  \sqrt{ \frac{E_{\rm{chem}} - E_{\rm{S}}}{E_{\rm{phys}} - E_{\rm{S}}} } \;\;\;  \right)^{-2} \left( \exp{\left(- \frac{E_{\rm{phys}} - E_{\rm{S}}}{E_{\rm{phys}} \text{+} T_{\rm{k}}} \right)}\right),
\label{eqn:prob-for-H-and-D-to-land-in-chemi-site}
\end{equation}
where $E_{\rm{chem}}$ and $E_{\rm{phys}}$ are respectively the chemisorption and physisorption binding energies, $E_{\rm{S}}$ is the energy of the saddle point between the two, and $T_{\rm{k}}$ is the kinetic temperature of the system. For high barriers between physisorbed and chemisorbed binding sites, physisorbed H atoms ("$\rm{H_{p}}$" below) with an oscillation factor of $\nu_{\rm{H_{p}}}$ will generally evaporate with a rate of $\beta_{\rm{H_p}}$, 
\begin{equation}
\beta_{\rm{H_{p}}} \; \text{=} \; \nu_{\rm{H_{p}}} \exp \frac{-E_{\rm{H_{p}}}}{T_{\rm{k}}}.
\label{eqn:H-physi-evap}
\end{equation}
By incorporating chemisorption into our models, H$_2$ may continue to be synthesised on the grains at higher temperatures than considered here. Because the different grain compositions are characterised by their variable-strength barriers against chemisorption, the efficiencies with which H$_2$ is synthesised on their surfaces are described by different equations. If the grain is carbonaceous and the barrier to transport from a physisorbed site to a chemisorbed site is low, then the H atoms may migrate from the former to the latter in order to produce H$_2$ with an efficiency of 
\begin{equation}
\epsilon_{\rm{H_{2}}}^{\rm{carb}} \; \text{=} \; \frac{1 - P_{\rm{H}}}{\left(1 \; \text{+} \; \frac{1}{4} \left(1 \; \text{+} \; \sqrt{\frac{E_{\rm{chem}} - E_{\rm{S}}}{E_{\rm{phys}} - E_{\rm{S}}}}\right)^2 \exp{- \frac{E_{\rm{S}}}{T_{\rm{k}}}} \right)}.
\label{eqn:H2-HD-form-carbonaceous-grains}
\end{equation}
Because the barrier between physisorbed and chemisorbed binding sites tends to be larger on silicate surfaces, it is consequentially more difficult for silicate-bound adatoms to migrate between the two. The efficiency of H$_2$ formation on this type of surface may be described, 
\begin{equation}
\begin{split}
\epsilon_{\rm{H_{2}}}^{\rm{sil}}& \; \text{=} \; \frac{1}{1\; \text{+} \; \frac{16T_{\rm{k}}}{E_{\rm{chem}} - E_{\rm{S}}}\exp \frac{-E_{\rm{phys}}}{T_{\rm{k}}} \exp \left(4 \times 10^9 a_{pc} \sqrt{E_{\rm{phys}} - E_{\rm{S}}} \right) }\\
& \; \text{+} \quad  2\frac{\exp{- \frac{E_{\rm{phys}} - E_{\rm{S}}}{E_{\rm{phys}} \; \text{+} \; T_{\rm{k}}}}}{\left(1 \; \text{+} \; \sqrt{  \frac{E_{\rm{chem}} - E_{\rm{S}}}{E_{\rm{phys}} - E_{\rm{S}}}}    \right)^2},\\
\end{split}
\label{eqn:H2-form-siliceous-grains}
\end{equation}
where $a_{pc}$ is the barrier width between physisorbed and chemisorbed sites, chosen to be 2.5 \r{A} here per \citet{barlow1976h2} and \citet{fromherz1993chemisorption}. Further details of Equations \ref{eqn:prob-for-H-and-D-to-land-in-chemi-site} -- \ref{eqn:H2-form-siliceous-grains} are described in \citet{cazaux2004h2} and \citet{cazaux2009hd}.

Applying the sticking coefficient and efficiency equations shown above, we write the final equation for the rate of H$_2$ formation on the different grain surfaces in our models as 
\begin{equation}
R_{\rm{s}} \big({\rm{H_2}}\big) \; \text{=} \; \frac{1}{2} n_{\rm{H}} \upsilon_{\rm{H}}S_{\rm{H}}\big(T_{\rm{k}}\big) \times \left( \left(n_{\rm{g}}\sigma \epsilon_{\rm{H_2}}\right)^{\rm{carb}} \; \text{+} \; \left(n_{\rm{g}}\sigma \epsilon_{\rm{H_2}}\right)^{\rm{sil}} \right),  
\label{eqn:modified-H2-formation}
\end{equation}
where \textit{n}$_{\rm{H}}$ is the number density of gas-phase H, $\upsilon _{\rm{H}}$ is the velocity of incoming H atoms which collide with the grain, \textit{n}$_{\rm{g}}$ is the number density of grains, and $\sigma$ is the cross-sectional surface area of the grain available for reaction.

\section{Results}
\label{sec:Results}

\subsection{Observational trends}
\label{sec:observational-agreement}
\begin{figure}
\centering
\includegraphics[width=8.5cm]{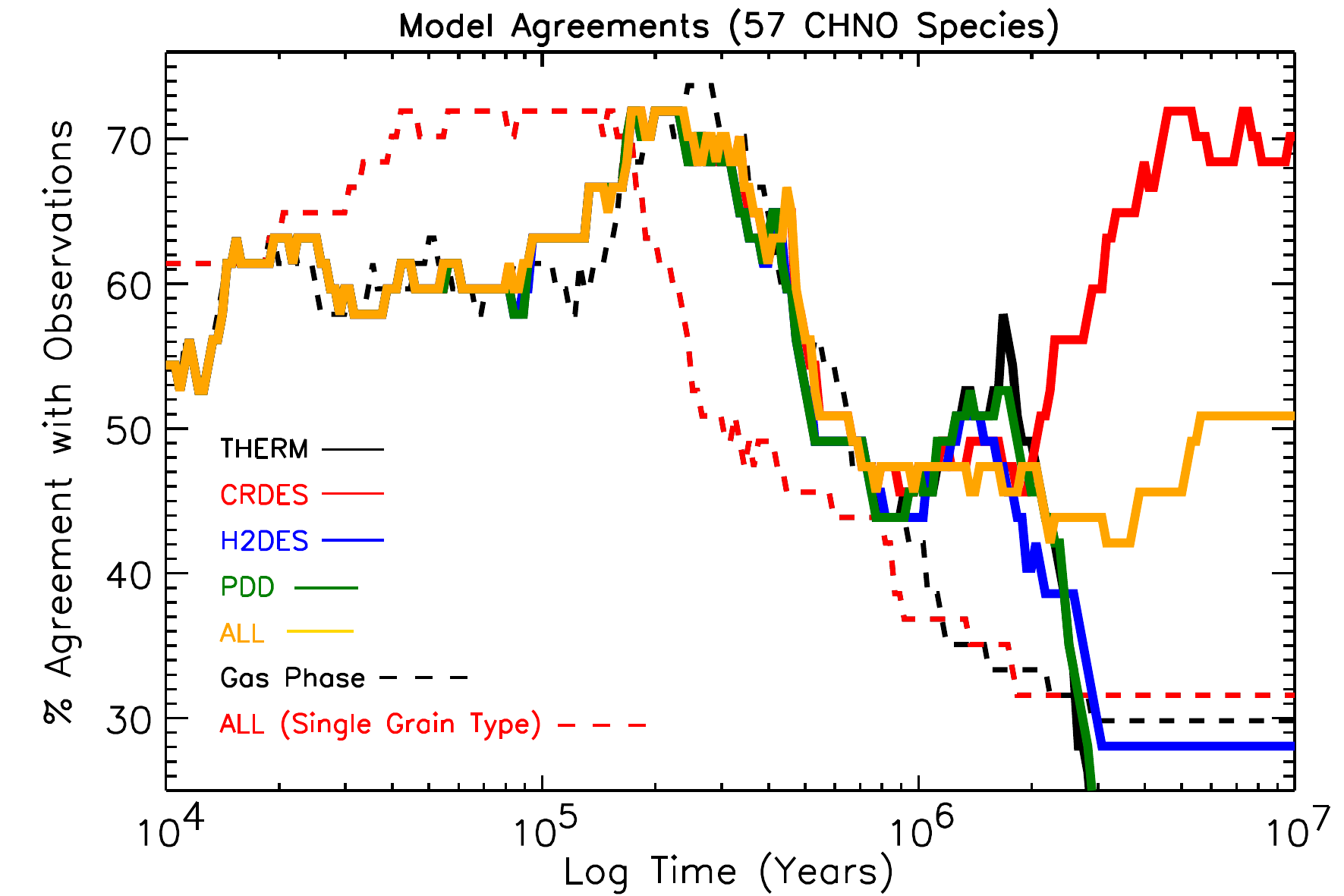}\\ 
\includegraphics[width=8.5cm]{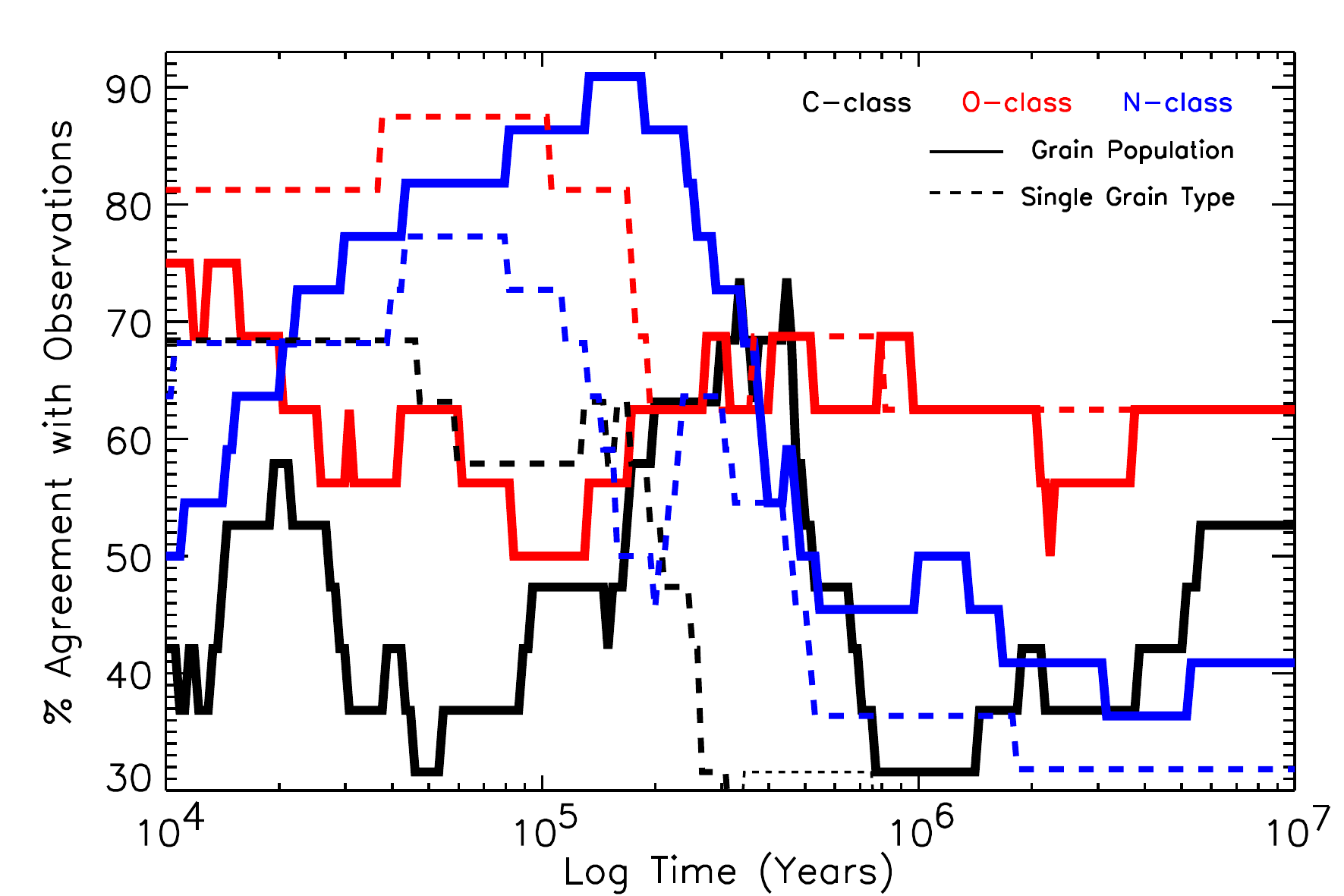}\\ 
\includegraphics[width=8.5cm]{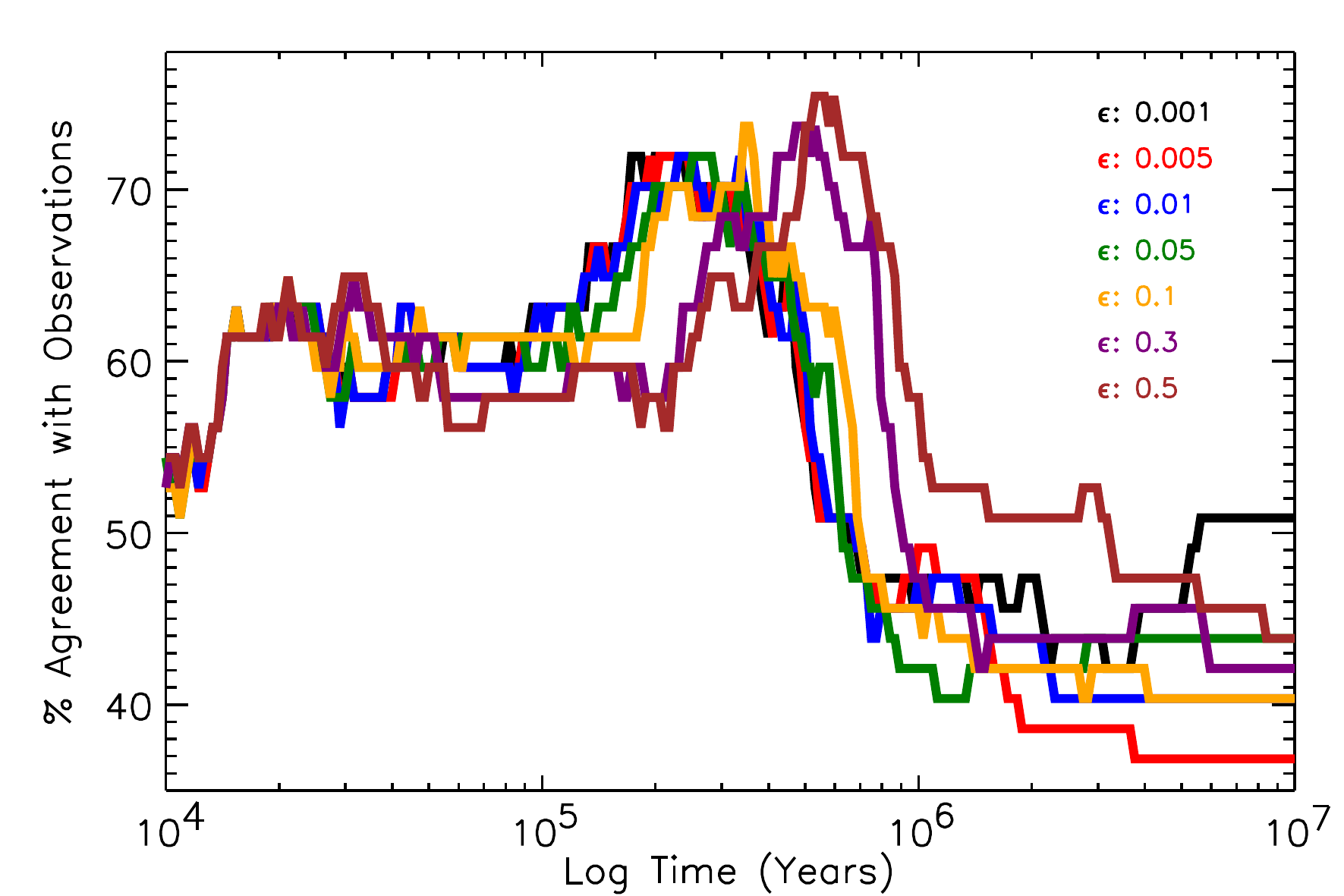} 
\caption{\emph{Top.} Agreement percentages in our ALL ($\epsilon$ = 0.001) model when the 57 atoms and molecules within the species classes bearing C, H, N, and/or O atoms observed towards TMC-1 (CP) are treated as a single group. \emph{Middle.} Percentage agreements of our ALL ($\epsilon$ = 0.001) model when the different "C-class," "O-class," and "N-class" groups are considered individually. Not pictured is an early-time agreement peak of the "C-class" group at times before 10$^2$ years. \emph{Bottom.} Agreements achieved by the 57 observed species in our ALL model when the H2DES $\epsilon$ parameter is varied.
              }
\label{SpeciesClassAgreements-CHNObearing}
\end{figure}

\def\starttable#1{%
  \renewcommand{\arraystretch}{0.95}%
  \minipage{0.99\textwidth}
      \centering
      \captionof{table}{#1}
      \tabular{lccc|lccc|lccc}
      \hline
      \hline
}
\def\stoptable#1{\\%
   \hline\endtabular\par\vspace{\abovecaptionskip}%
   {\footnotesize #1}%
   \endminipage}
\def\R #1|#2|#3|#4{ #1&#2&#3&#4}

\begin{table*}
\centering
\starttable{Fractional abundances (listed with respect to the total H nucleon number density, H$^{-1}$) of species observed towards the TMC-1 Cyanopolyyne Peak which we have used for our agreement analysis. "LLIM" denotes the lower limit; "ULIM" denotes the upper limit (or maximum possible limit); and the notation x(y) represents a number of the form x $\times$ 10$^{y}$. }
Species           				&X$_{10}$  	&X$_{10}$	& Abund		& Species				&X$_{10}$  	&X$_{10}$ 	& Abund  		& Species    			&X$_{10}$  	&X$_{10}$ 	& Abund  	\\ 
                        				&(LLIM)     & (ULIM)   		& Refs 		&               			&(LLIM)    		& (ULIM)   	& Refs     		&             				&(LLIM)    		& (ULIM)    	& Refs      \\  
\hline	                                                 	
CO						&4.00E-06	&7.30E-05	&2, 18		&C					&>5.00E-07	&--			&8			&O$_2$				&--			&<3.85E-07	&18\\
OH						&1.24E-09	&1.30E-07	&16			&H$_2$O				&--			&<3.50E-07	&7			&C$_2$H				&1.90E-09	&4.60E-07	&15\\
C$_2$					&2.50E-09	&2.50E-07	&2, 18		&NO					&1.50E-09	&1.50E-07	&18			&NH$_3$				&1.17E-10	&1.02E-07	&4, 19\\
HNC						&1.00E-09	&1.00E-07	&18			&H$_2$CO			&1.00E-09	&1.00E-07	&2			&HCN				&1.00E-09	&1.00E-07	&18\\
HC$_3$N					&9.75E-10	&9.75E-08	&17			&CH$_3$CCH			&3.15E-10	&8.30E-08	&4, 19		&CH					&4.97E-10	&6.53E-08	&16\\
HCO$^{\text{+}}$			&4.00E-10	&4.00E-08	&18			&C$_3$H$_2$			&2.90E-10	&2.90E-08	&9			&HC$_5$N			&2.10E-10	&3.50E-08	&10\\
C$_4$H					&2.70E-10	&2.70E-08	&20			&CN					&2.50E-10	&2.50E-08	&18			&CH$_2$CHCN		&2.25E-10	&2.25E-08	&18\\
CH$_2$CN				&2.50E-10	&2.50E-08	&18			&CH$_3$CHCH$_2$	&2.00E-10	&2.00E-08	&14			&CH$_2$NH			&--			&<1.77E-08	&10\\
H$_3$CO$^{\text{+}}$		&--			&<1.55E-08	&18			&HCNH$^{\text{+}}$		&1.00E-10	&1.00E-08	&18			&HC$_7$N			&5.00E-11		&7.00E-09	&10\\
C$_3$H					&4.15E-12	&4.25E-10	&9			&HC$_9$N			&1.25E-11		&1.45E-09	&10			&CH$_3$CN			&3.00E-11		&3.00E-09	&18\\
CH$_2$CO				&3.00E-11		&3.00E-09	&18			&C$_3$N				&3.00E-11		&3.00E-09	&18			&C$_6$H				&2.81E-11		&2.97E-09	&17\\
C$_5$H					&2.90E-11		&2.90E-09	&6			&HNCO				&2.05E-11		&2.05E-09	&18			&N$_2$H$^{\text{+}}$	&2.00E-11		&2.00E-09	&18\\
CH$_3$C$_6$H			&2.10E-11		&4.10E-09	&12			&CH$_3$OH			&1.17E-11		&1.29E-09	&4, 19		&H$_2$CCC			&1.05E-11		&1.05E-09	&9\\
HCOOH					&1.00E-11		&1.00E-09	&18			&CH$_3$C$_4$H		&8.86E-11		&9.54E-09	&12			&C$_3$O				&5.00E-12	&5.00E-10	&18\\
HC$_3$NH$^{\text{+}}$		&5.74E-13	&1.04E-10	&4, 19		&CH$_3$C$_5$N		&3.50E-12	&3.90E-10	&13			&C$_3$N$^{-}$			&--			&<3.50E-10	&18\\
C$_2$O					&3.00E-12	&3.00E-10	&18			&CH$_3$CHO			&1.26E-12	&2.75E-10	&4, 19		&C$_8$H				&2.30E-12	&2.30E-10	&18\\
CH$_3$C$_3$N			&2.05E-12	&2.45E-10	&11			&HNC$_3$			&1.60E-12	&2.20E-10	&1			&C$_5$N				&1.50E-12	&1.50E-10	&5\\
H$_2$CN					&7.50E-13	&7.50E-11		&3			&C$_7$H				&--			&<7.50E-11	&6			&C$_6$H$^{-}$			&6.45E-13	&8.25E-11		&17\\
C$_8$H$^{-}$				&1.05E-13	&1.05E-11		&18			&HCNO				&--			&<6.50E-12	&18			&C$_4$H$^{-}$			&2.80E-14	&5.20E-12	&17\\
\label{tab:DC-Observed-Abunds}
\stoptable{(1)~\citet{kawaguchi1992detection}; (2) \citet{ohishi1992molecular}; (3) \citet{ohishi1994detection}; (4) \citet{ohishi1998chemical}; (5) \citet{guelin1998}; (6) \citet{bell1999observations}; (7) \citet{snell2000water}; (8) \citet{charnley2001gasgrainC} and references therein; (9) \citet{fosse2001molecular}; (10) \citet{kalenskii20044}; (11) \citet{lovas2006hyperfine}; (12) \citet{remijan2006methyltriacetylene}; (13) \citet{snyder2006confirmation}; (14) \citet{marcelino2007discovery}; (15)  \citet{sakai2010abundance}; (16) \citet{suutarinen2011ch}; (17)  \citet{cordiner2013ubiquity}; (18) \citet{mcelroy2013umist} and references therein; (19) \citet{gratier2016new}; (20)  \citet{oyama2020reevaluation}}
\hspace{10pt}%
\end{table*}

\begin{figure*}
\centering
\includegraphics[width=16.4cm,clip]{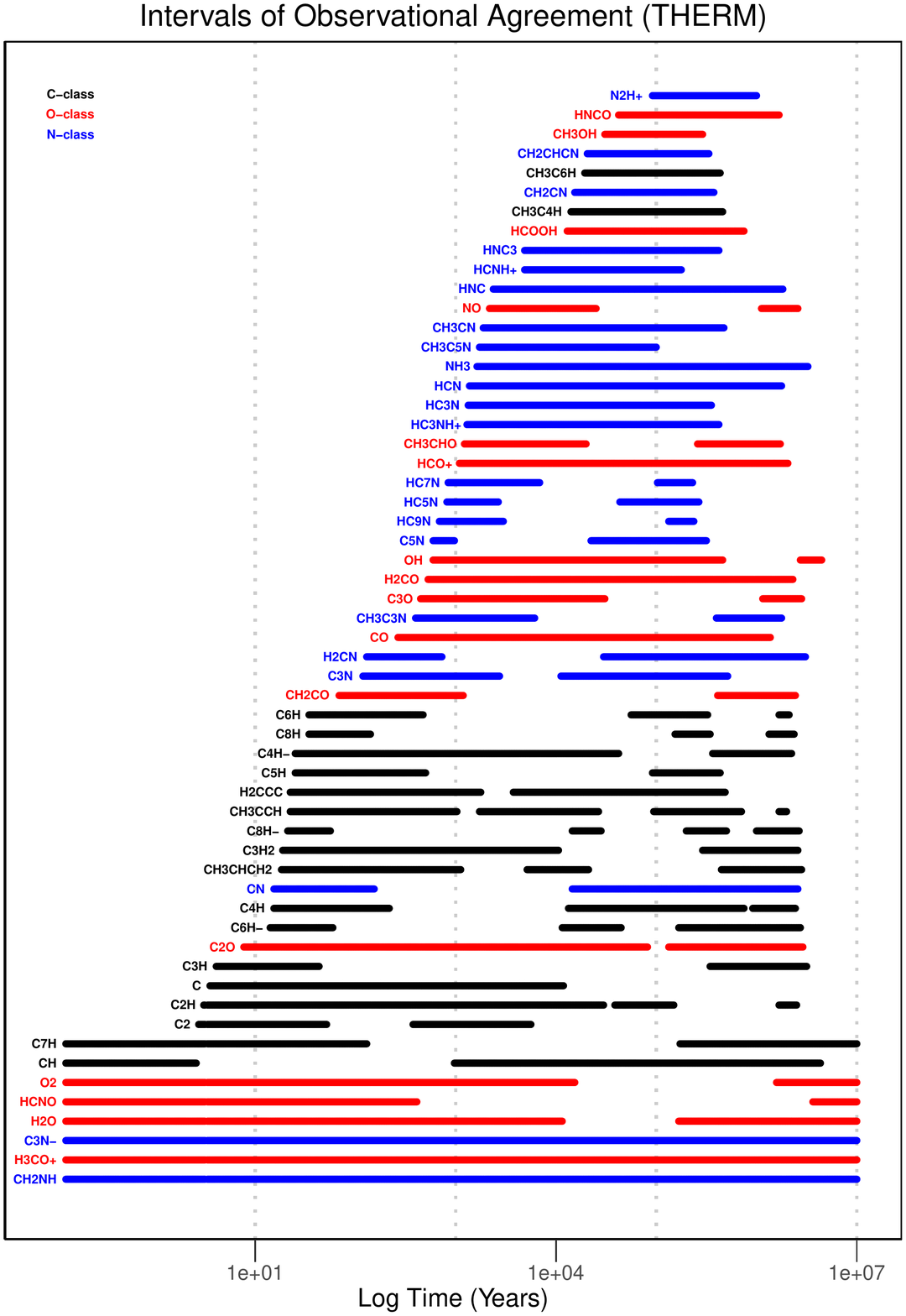}
\caption{Duration of agreement for the 57 atomic and molecular species which have been previously observed towards TMC-1 (CP) as calculated by our THERM model. }
\label{AgreementTimeIntervalsTHERM}
\end{figure*}

\begin{figure*}
\centering
\includegraphics[width=16.4cm,clip]{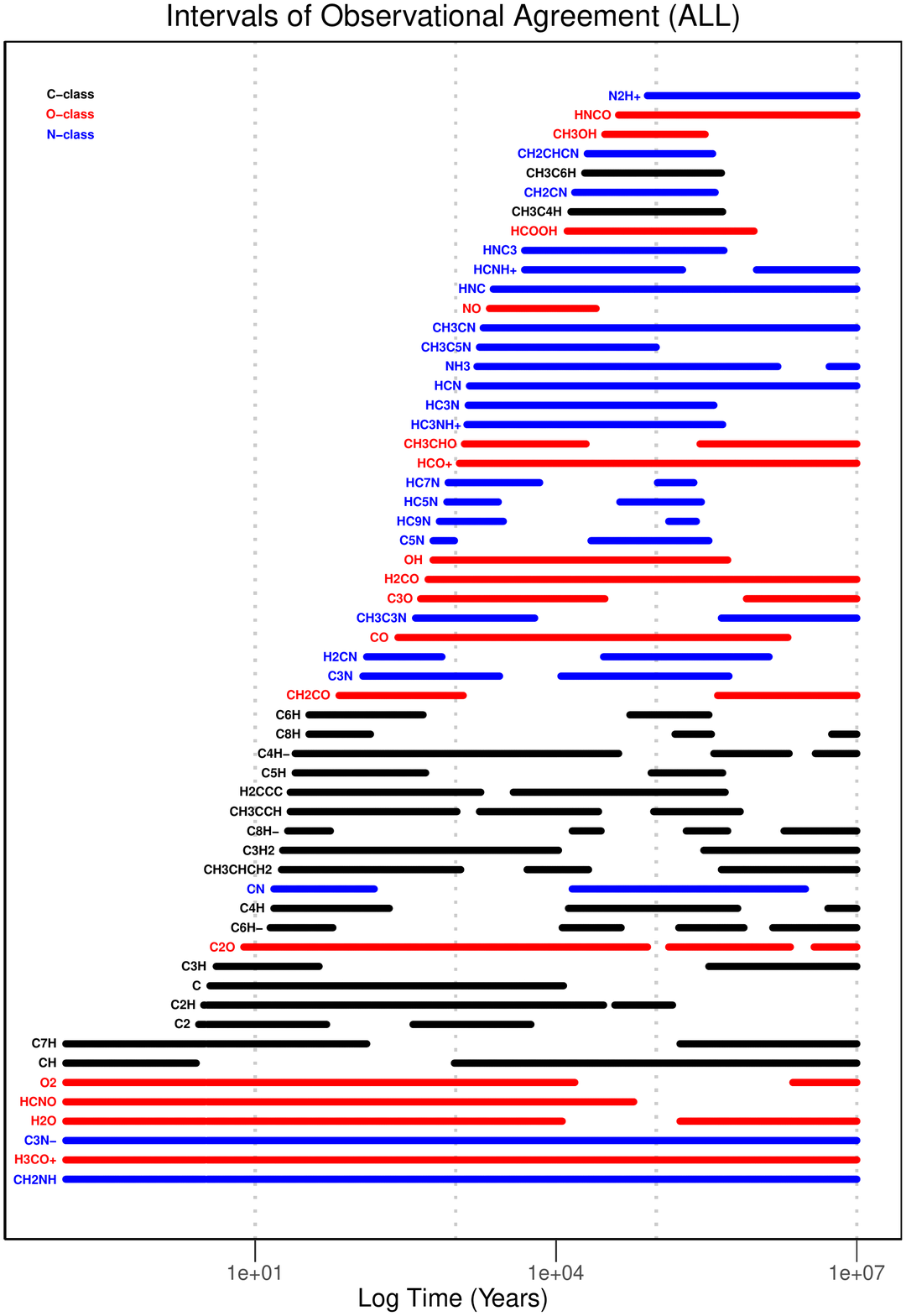}
\caption{Same description as in Fig. \ref{AgreementTimeIntervalsTHERM} but for the ALL  ($\epsilon$ = 0.001) model. The inclusion of non-thermal desorption extends the range of "good agreement" for many of the modelled species.}
\label{AgreementTimeIntervalsALL}
\end{figure*}

To assess how well our dark cloud models represent the actual chemistry of low-mass star-forming environments, we compare our calculated time-dependent abundances with the abundances of species which have been detected towards TMC-1 (CP) (Fig. \ref{SpeciesClassAgreements-CHNObearing}; Table \ref{tab:DC-Observed-Abunds}). In light of \citet{dobashi2018spectral, dobashi2019discovery}'s detection of likely physical substructures within TMC-1 (CP) and \citet{choi2017dynamical}'s proposal that the region is likely to be in the process of rapid core formation due to the effects of colliding molecular clouds, we consider the observational agreement of our models over the course of the chemical evolution under a variety of scenarios: when 1.) the 57 observed CHNO species (species which consist solely of C, H, N, and/or O atoms) in our different desorption models are calculated as a single group, 2.) the species are divided into different subgroups in our ALL ($\epsilon$ = 0.001) model, and 3.) when the efficiency parameter $\epsilon$ is varied from $\epsilon$ = 0.001 (conservative) to $\epsilon$ = 0.5 (highly efficient) in our ALL model. While we have included S-chemistry in our calculations, we exclude S-bearing species from our analysis due to the inherent uncertainties of the elemental S abundance as well as in the sulphur reaction network itself, both of which make it difficult to fit the observations of S-bearing molecules in a comprehensive manner. We note that \citet{vidal2017sulphur} and \citet{laas2019sulfur} have produced new models of sulphur chemistry based on significant extensions of current reaction networks and the physical evolution of molecular clouds, respectively. 


In order to obtain the fractional abundance estimates listed in Table \ref{tab:DC-Observed-Abunds}, we have used a molecular hydrogen column density of N(H$_2$) = 10$^{22}$ cm$^{-2}$ together with column densities derived from observational data. It should be noted that the derivation of observed column densities can be made under a variety of assumptions, from optically thin emission and assumed rotational temperatures, to the use of hyperfine intensities to derive optical depths, to full scale radiative transfer calculations and the use of multiple line transitions to derive optical depths and excitation temperatures. One should note that while the latter approach may give the most accurate column densities, many of the complex species detected in TMC-1 have no published collisional cross sections. Thus, observationally-derived column densities can vary by a factor of a few depending on the technique adopted. Our use of upper and lower limits on the column densities mitigates somewhat against these uncertainties. Given the inherent uncertainties in the modelling as well (e.g., errors in the rate equations, unknown product branching ratios, unknown parameter values for the physical processes which affect the gas-grain chemistry, etc.), we have taken the default approach of considering our modelled values to be within "good agreement" of observations where they fall within an order of magnitude of their reported values, considering the uncertainties in the observations. For those instances in which the ranges of the Bayesian statistical method uncertainty margins reported by \citet{gratier2016new} to account for known outliers in the observations exceed these limits, however, we adopt those instead, unless the values produced by \citet{gratier2016new} are suggestive of an upper limit for molecules which have clear detections.




When we consider the observed CHNO species as a single group (Fig. \ref{SpeciesClassAgreements-CHNObearing}, top panel), the maximum agreement ($\sim$72\%) achieved by our ALL model when calculating the solid-state formation of H$_2$ on a single grain type occurs between $\sim$4 $\times$ 10$^4$ and 1.5 $\times$ 10$^5$ years. When the process occurs instead across our chosen population of realistic grain types, however, the interval of maximum agreement (also $\sim$72\%)\footnote{74\% in our "Gas Phase" model.} is shifted to later times, reaching its highest value around $\sim$2 $\times$ 10$^5$ years. From around 2 $\times$ 10$^5$ years and through the remainder of the modelled evolution, the general agreement achieved by the ALL model remains, to varying degrees, better when adopting the updated treatment for H$_2$ formation than it does when defaulting to the standard method. The similarity across our results before $\sim$2 $\times$ 10$^{5}$ years in the models which include a population of different grain types is due to the fact that during this stage of the chemical evolution, the overarching chemistry is still largely dominated by gas-phase processes. While the general agreement trends for these models are similar until around 10$^{6}$ years, they deviate progressively as the freeze-out ensues. This is because the gas-grain mechanisms which distinguish our different desorption models from each other are generally long-term effects and therefore typically do not become manifest in the calculated results until later years, when much of the atomic and molecular material has started to accrete onto the grains. For times beyond $\sim$2 $\times$ 10$^6$ years, the CRDES model produces the best agreement with observations ($\sim$72\% at 5 $\times$ 10$^6$ years), followed by the ALL model (51\% for times greater than $\sim$5 $\times$ 10$^6$ years). The improved late-time agreement of C-bearing species in the CRDES model with respect to the other models is largely the result of an enhanced presence of atomic C in the gas phase of the CRDES model at late times. This leads to the enhanced production of CH$^{\text{+}}$ through proton exchange reactions between C and H$_3^{\text{+}}$ and thus the enhanced late-time production of the larger hydrocarbons thereby derived. Our models suggest that this effect is particularly enhanced when the formation of H$_2$ occurs across a population of different grain types, as evidenced by the improved agreement in the CRDES models at late times. Interestingly, our results also suggest that the general agreement at times before $\sim$10$^6$ years is more sensitive to the mechanisms employed for the formation of H$_2$ on the grains than it is to the presence of grains in the first place. 





As different regions of the cloud may be subjected to different chemical and physical conditions, we wish to ascertain whether or not different groups of species within our models are better representative of observations at different times. To identify the different groups, we consider the heaviest atom contained in each of the 57 individual CHNO species. In this scheme, each species is assigned to one and only one "class." For example, molecules in the "C-class" (19 species) consist of only C and H atoms; molecules in the "N-class" (22 species) must contain an N atom (but can also have C atoms); and molecules in the "O-class" (16 species) must contain an O atom (but can also have C and N atoms). In Figs. \ref{AgreementTimeIntervalsTHERM} and \ref{AgreementTimeIntervalsALL}, the species within the different C-class, O-class, and N-class groups are denoted, respectively, by the colours black, red, and blue. In the middle panel of Fig. \ref{SpeciesClassAgreements-CHNObearing}, we present the time-dependent agreements for these different groups in our ALL ($\epsilon$ = 0.001) model when a realistic population of grains is assumed for the formation of H$_2$ versus when the process occurs on a single, arbitrary grain type. When a population of grains is assumed, the peak agreement of the C-class (85\%) occurs at very early times ($\sim$(3 -- 5) $\times$ 10$^1$ years), that for the N-class (91\%) occurs between (1 -- 2) $\times$ 10$^5$ years, and that for the O-class (75\%) spans from $\sim$2 $\times$ 10$^{3-4}$ years. The C-class and O-class then re-achieve agreements of 74\% and 69\%, respectively, between $\sim$(3 -- 5) $\times$ 10$^5$ years while the agreement of the N-class simultaneously decreases from $\sim$73\% to $\sim$60\%. In contrast, when only an arbitrary grain type is considered, the maximum agreements of the C-class and N-class species groups are reduced to 79\% (also at early times) and 77\%, respectively, whereas that of the O-class species group increases to 88\%. The confluence of curves around 3 $\times$ 10$^5$ years in our models which adopt a population of different grain types is in general agreement with the range of ages expected for TMC-1 (CP) based on its dynamics and observed chemistry. While this method of categorisation is somewhat coarse,\footnote{For example, HCNO and HNCO could also be grouped with either the N-class or the C-class.} our results at least demonstrate in a broad sense that the maximum agreement achieved is higher when the different chemical subgroups are treated separately than it is when they are treated as a collective.



In the bottom panel of Fig. \ref{SpeciesClassAgreements-CHNObearing}, we illustrate the effect that varying the $\epsilon$ value has on the agreement achieved by the ALL model. While the higher values of $\epsilon$ generally improve the modelled agreement at late years, they also tend to suppress the agreement between $\sim$5 $\times$ 10$^4$ and $\sim$3 $\times$ 10$^5$ years. The best agreement of these models for times later than 10$^4$ years (75\%) occurs around (5 -- 6) $\times$ 10$^5$ years in the $\epsilon$ = 0.5 model. The variance of our results for different $\epsilon$ values suggests that the efficiency of the H2DES process is an important parameter which can affect the course of the gas-grain chemistry throughout the evolution of the cloud. These results are in agreement with previous studies wherein the desorption of mantle material through H$_2$ dissociation, and via chemical desorption in general, was found to be important \citep[e.g.,][]{roberts2007desorption, hocuk2015interplay, minissale2016hydrogenation}.

In Figs. \ref{AgreementTimeIntervalsTHERM} and \ref{AgreementTimeIntervalsALL}, we show the intervals of "good agreement" in our THERM and ALL models for each of the 57 different CHNO species observed towards TMC-1 (CP). In general, we find that the intervals of "good agreement" for the C-class species occur at times before those of the N-class and O-class species. However, there are exceptions, such as the early-time agreements of CN and C$_2$O and the late-time agreements of CH$_3$C$_4$H and CH$_{3}$C$_6$H. The prolonged agreements of the 8 species at the bottom of these plots is a result of the fact that their reported abundances are known only as either upper or lower limits.


The general trend that we see in our agreement plots is that when we distinguish the observed species by class, 1.) the agreement percentages themselves are improved, and 2.) the agreement itself is time-dependent, depending on the class. While the chemistry before $\sim$10$^2$ years is dominated by CN and the lighter C-class species, the chemistry between $\sim$10$^3$ and 10$^5$ years is instead dominated by N-class and O-class species. When all 57 species are treated as a single group, the interval of "best agreement" occurs around (2 -- 3) $\times$ 10$^{5}$ years, when the N-class and O-class species are still abundant in the gas phase and many of the C-class species have re-achieved their agreement thresholds. Beyond $\sim$2 $\times$ 10$^5$ years, the chemistry is characterised by the effectiveness of the gas-grain processes employed as well as gas-phase routes which are driven by the behaviour of the desorbed species, notably atomic C.





The fact that different species types in our models achieve their best agreements with observations at different times could reasonably be explained by physical factors which influence the substructures within the Cyanopolyyne Peak. First of all, the rate at which the gas-phase chemistry proceeds is roughly proportional to the density squared; and the rate of accretion, in turn, is proportional to the density. Therefore, neighbouring clumps of material which possess different densities could evolve on slightly different time-scales, even if their chemistries start at the same time. Secondly, as discussed in Section \ref{sec:TMC-1}, there is observational evidence for both turbulence as well as multiple velocity components within the core. Given the findings of \citet{choi2017dynamical} and \citet{dobashi2018spectral} which support the theory that TMC-1 (CP) is in the early stages of a collision-induced self-gravitational collapse, it is plausible that as-yet unresolved clumps of material within the aggregate might collide with each other and effectively "reset" the chemistry through the re-injection of mantle material into the gas phase. In our models, for example, the CH$_4$, C$_2$H$_2$, and C$_3$H$_2$, and CH$_3$CCH ices have abundances greater than 10$^{-7}$ H$^{-1}$ at times earlier than 10$^5$ years; their return to the gas through such physical processes could conceivably re-initiate the "early-time" hydrocarbon chemistry in localised regions of the cloud.

\subsection{NH$_3$ chemistry}
\label{sec:NH3}

\begin{figure*}
\centering
\includegraphics[angle=0,width=85mm]{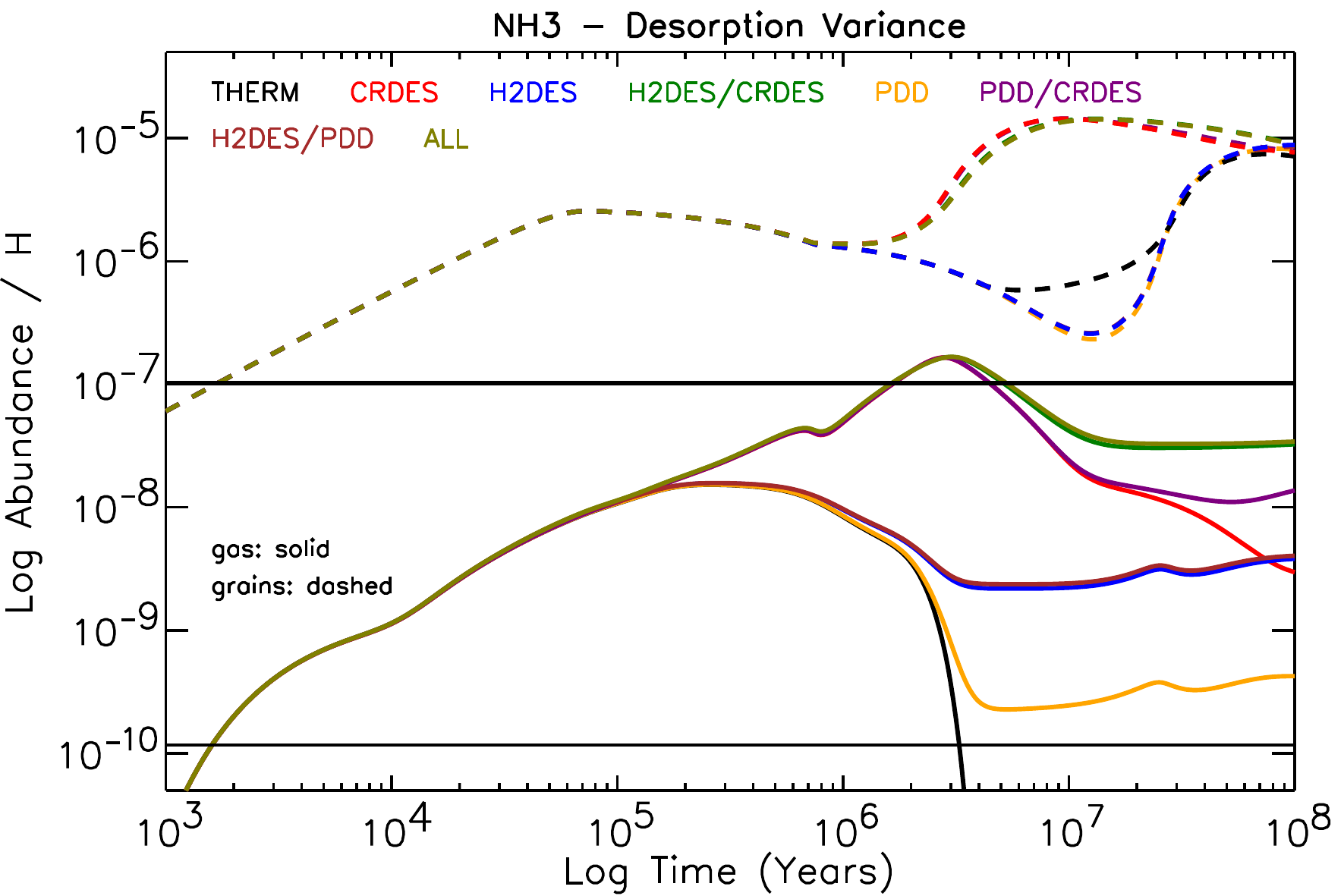}
\includegraphics[angle=0,width=85mm]{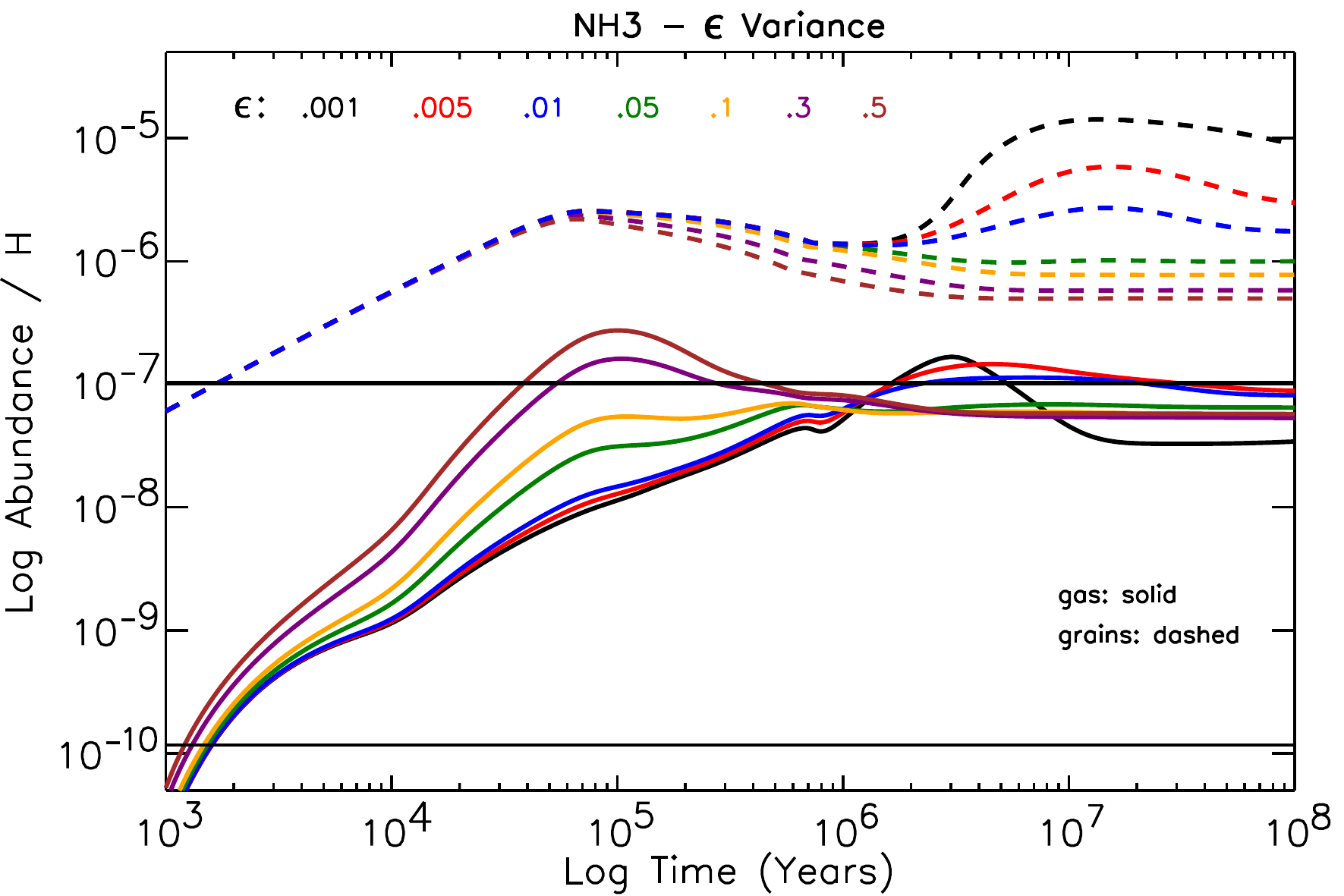}
  \caption[Model description as in Fig. \ref{fig:desorptions-NH3-HC3N}.]{\small{Time-dependent abundances of NH$_{3}$ and NH$_{3,\rm{ice}}$ in our different models when the desorption mechanisms are varied (left) and in the ALL model when the value of $\epsilon$ is varied (right). Horizontal black lines indicate the margins of "good agreement" with observations. "Good agreement" for gaseous NH$_{3}$ is achieved by 2 $\times$ 10$^3$ years, and high abundances are maintained throughout the evolution when non-thermal desorption is efficient. The NH$_{3, {\rm{ice}}}$ abundances in the H2DES/PDD model are overlapped by those in the H2DES model, and the NH$_{3,{\rm{ice}}}$ abundances in the PDD/CRDES model are overlapped by those in the CRDES model.}}
  \label{fig:NH3-GNH3-epsilonVary}
\end{figure*}


As we see in Fig. \ref{fig:NH3-GNH3-epsilonVary} (left panel), the calculated gas-phase abundances of NH$_3$ in our gas-grain models which consider a realistic population of different grain types reach our metric for "good agreement" at times as early as 2 $\times$ 10$^{3}$ years within the large observational uncertainties toward TMC-1 (CP) as determined by \citet{gratier2016new}. By 5 $\times$ 10$^4$ years, our values reach that reported by \citet{feher2016structure} of 7 $\times$ 10$^{-9}$ H$^{-1}$; and at late times, the value of the NH$_3$ abundance can vary by nearly 3 orders of magnitude, depending on the mechanisms chosen for non-thermal desorption. Comparing their gas-phase results when incorporating the KIDA versus UDfA RATE12 networks, \citet[][their Fig. 4]{agundez2013chemistry} found that for C-rich conditions, the UDfA model produces more NH$_3$ at early times than does the KIDA model (by a factor of $\sim$7 at 10$^4$ years) and that the KIDA model overproduces the NH$_3$ abundance for times later than $\sim$3 $\times$ 10$^5$ years. When evaluated against the \citet{gratier2016new} limits, their UDfA model remains in good agreement with observations for the remainder of the evolution. When we inspect the behaviours of the gas-grain models herein for different desorption combinations, we find that the simultaneous inclusion of both CRDES and H2DES has a synergistic effect on the quantity of NH$_{3}$ produced in the gas phase at the latest times (e.g., $\sim$5 $\times$ 10$^{-9}$ H$^{-1}$ and 1 $\times$ 10$^{-8}$ H$^{-1}$ respectively at 2 $\times$ 10$^7$ years in the H2DES and CRDES models versus 3 $\times$ 10$^{-8}$ H$^{-1}$ at this time in the H2DES/CRDES and ALL models). While the abundance of gas-phase NH$_3$ tends to decrease at late years in our CRDES model, the inclusion of PDD to the chemistry stabilises the abundance from 10$^7$ years onwards, as seen in the PDD/CRDES model. In the gas phase, the abundances in the CRDES, H2DES, and PDD models begin to deviate from those in the THERM model around 2 $\times$ 10$^5$,  7 $\times$ 10$^5$ and 2 $\times$ 10$^6$ years, respectively. Our models suggest that while the NH$_3$ abundance remains in good agreement with observations at late times when the species cycling is dominated by H2DES and/or PDD, it is in fact overproduced from around (2 -- 5) $\times$ 10$^6$ years when CRDES dominates. Because the cycling of N$_2$ between the gaseous and solid states in our models is most efficient at later times ($t$ $\gtrsim$ 10$^{6}$ years) when cosmic-ray heating is relevant, the abundances of N-bearing species such as NH$_3$ which are derived from N$_2$ are generally highest in our models at the same time.

When all of the desorption mechanisms are considered simultaneously and $\epsilon$ \text{=} 0.001, the gaseous and solid-state NH$_3$ abundances tend to be dominated by CRDES. However, when we vary $\epsilon$ to higher values (Fig. \ref{fig:NH3-GNH3-epsilonVary}, right panel), we see that 1.) the peak abundance of NH$_3$ is transferred to earlier times in the gas phase,\footnote{$\sim$10$^5$ years for $\epsilon$ \text{=} 0.3 and 0.5 versus $\sim$3 $\times$ 10$^6$ years for $\epsilon$ \text{=} 0.001.} and 2.) the late-time abundance of NH$_{3,\rm{ice}}$ decreases by $\sim$1.5 orders of magnitude due to the fact that the molecule is more readily removed from the grains by the H2DES process. As we saw in the left panel of Fig. \ref{fig:NH3-GNH3-epsilonVary}, the abundance of NH$_{3,\rm{ice}}$ on the grains is further reduced when CRDES is excluded from the calculations entirely. We find that the enhancement of NH$_{3,\rm{ice}}$ at late times in our models begins when $\epsilon$ is between 0.01 and 0.05 and increases as the value of $\epsilon$ decreases.

\begin{figure}
   \resizebox{\hsize}{!}{\includegraphics{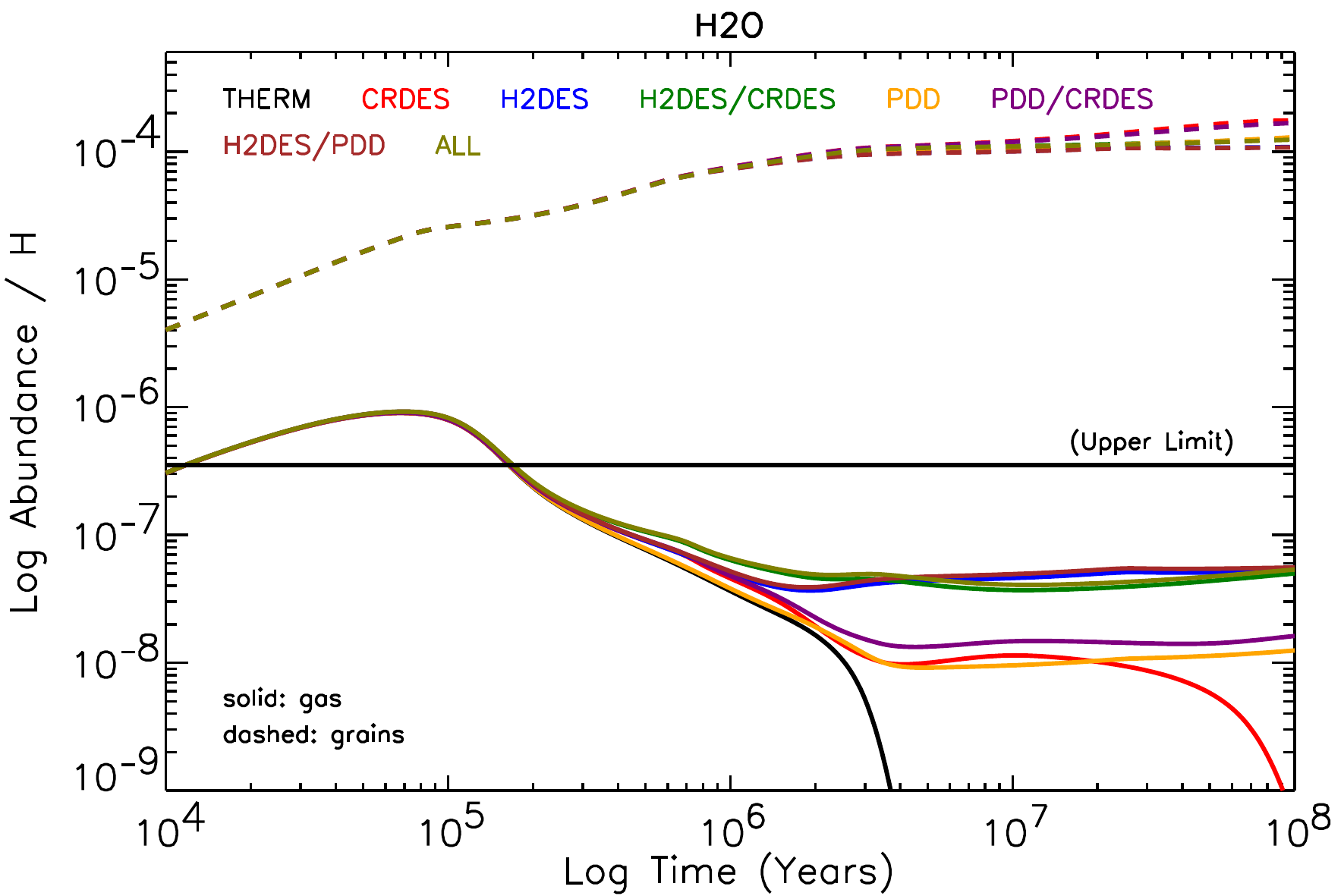}}  
   \resizebox{\hsize}{!}{\includegraphics{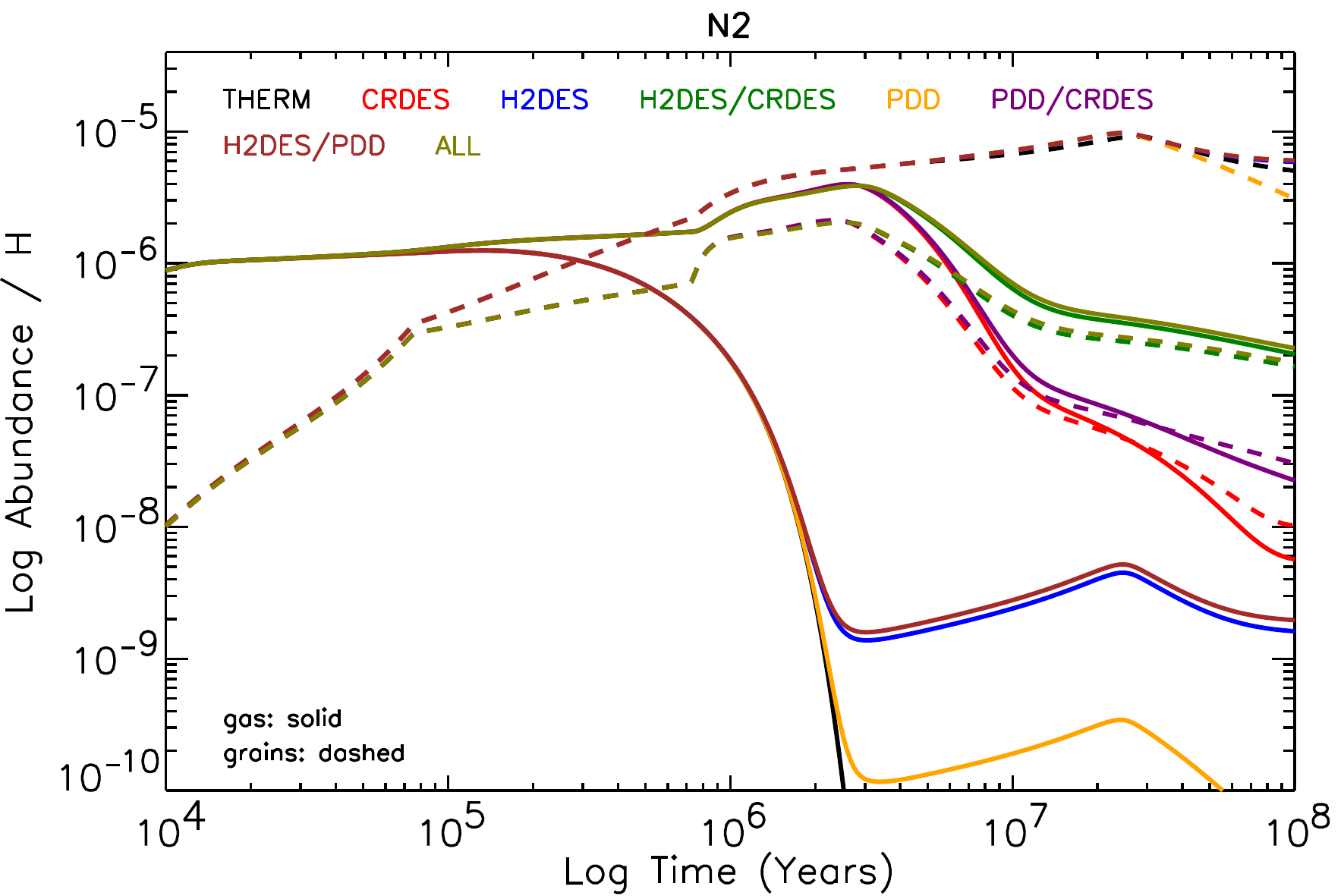}}    
   \resizebox{\hsize}{!}{\includegraphics{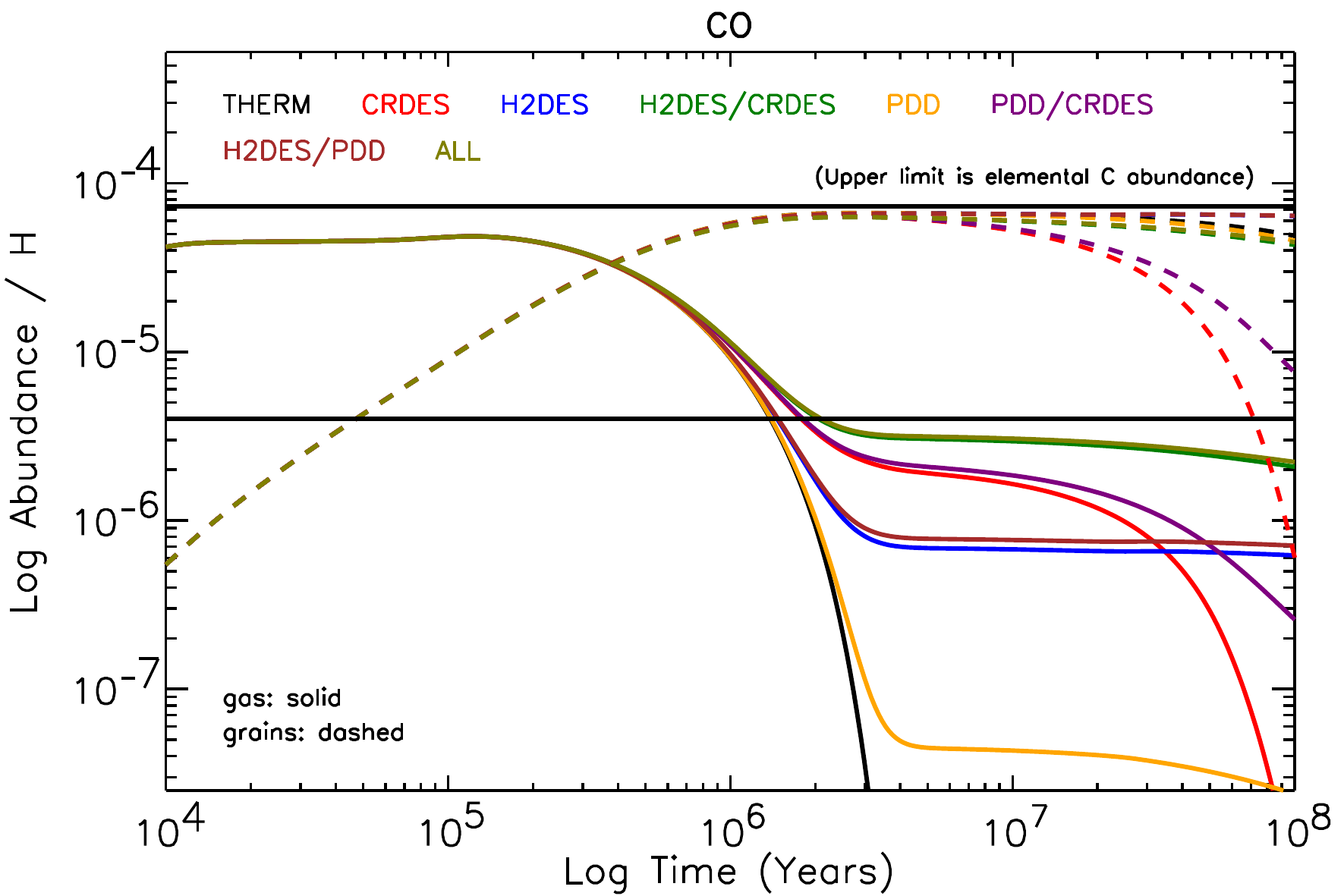}}    
   \resizebox{\hsize}{!}{\includegraphics{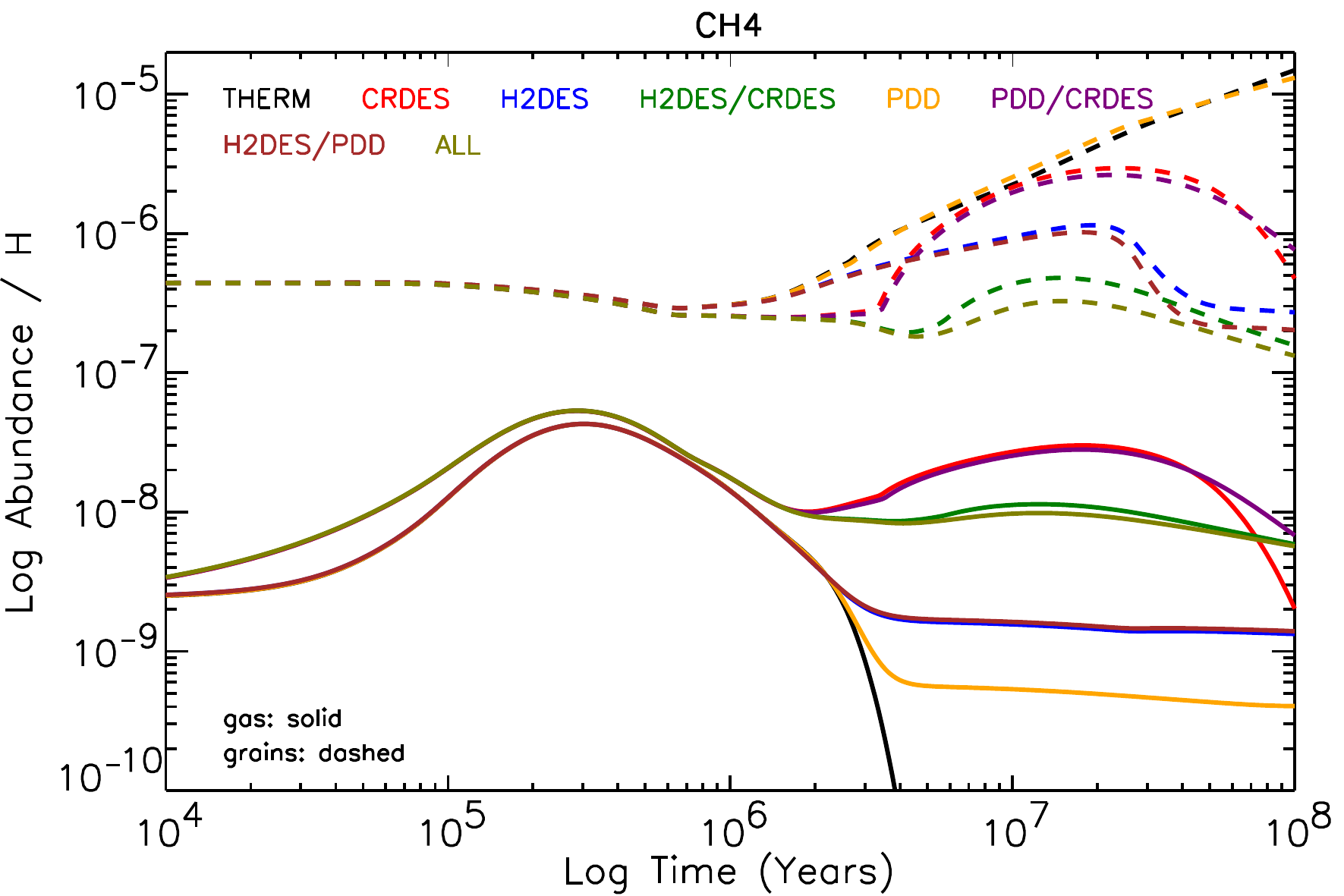}}    
  \caption[Model description as in Fig. \ref{fig:desorptions-NH3-HC3N}.]{\small{Time-dependent abundances of gaseous and solid-state H$_2$O, N$_2$, CO, and CH$_4$  in our different desorption models. These are the species whose abundances dominate the re-injection of O, C, and N atoms into the gas phase through gas-grain species cycling.}}
  \label{fig:H2O-GH2O-N2-GN2-CO-GCO}
\end{figure}




In general, the formation of NH$_3$ is thought to proceed in dark-cloud environments through the rapid hydrogenation of N atoms on the grains and N$^{\text{+}}$ ions in the gas phase \citep[for a more thorough review of the N chemistry, see, e.g.,][]{herbst2008chemistry, hilyblant2010nitrogen, legal2014interstellar, fedoseev2015low}. To derive N and N$^{\text{+}}$ in the first place, CN is thought to collide with N to form N$_2$, and N$_2$ then reacts with He$^{\text{+}}$ to yield N$^{\text{+}}$, N, and He. In our models, the production of NH$_3$ is enhanced at early times in the gas phase due to the efficient production of CN caused by the early-time enhancement of C$_{\rm{n}}$H species resulting from an efficient early-time network of hydrocarbon reactions. Since the early-time ($t$ $\lesssim$ 10$^3$ years) chemistry is highly dependent on the initial conditions adopted for the models, however, we focus our investigation on the chemistry from 10$^4$ years onwards, when the cloud has presumably been sufficiently processed that the chemistry has "forgotten" the details of its initial conditions.\footnote{For example, the efficient production of hydrocarbons in our models at very early times is largely the result of our having initialised the chemistry with C$^{\text{+}}$ rather than neutral C. Due to the necessarily higher ionisation accompanying this choice, our models would naturally be expected to demonstrate a richer early-time hydrocarbon chemistry than would, say, a model that begins with all of the atoms in a neutral state.} From around 10$^{4}$ years in our models, the production of NH$_3$ originates with the dissociation of NO by He$^{\text{+}}$ atoms. The general synthesis of NO at this time is related to that of water chemistry since the dissociative recombination of H$_3$O$^{\text{+}}$ yields OH, which then reacts with an N atom to form NO \citep[as discussed in, e.g.,][]{hilyblant2010nitrogen}. At later times, NO is still formed by this gas-phase route but where the water is provided by the desorption of water ice.

While the dominating desorption process for N$_2$ and CO in our models is CRDES, H$_2$O is instead most efficiently removed from the grains by H2DES (due to the fact that the H2DES process does not depend on the binding energies of species in their binding sites, a value which is high for H$_2$O). As noted above with the NH$_3$ abundances, the combination of H2DES and CRDES has a synergistic effect on the late-time N$_2$ (and hence NO) abundances in the gas phase of our $\epsilon$ = 0.001 ALL model. Indeed, it is the enhancement of N$_2$ which allows for the enhancement of both NO and NH$_3$. Because the cycling process allows N$_2$ to remain prevalent in the gas at late times, the abundances of N-bearing species derived from it -- for example, NH$_3$, NO, CH$_3$CN, NH$_2$CN, HCN, HNC, and the C$_{\rm{n}}$N (n = 1 -- 4) chains -- can continue to be produced in the gas phase at late times as well, as long as the cycling of N$_2$ is efficient. Time-dependent abundances of the dominant reservoirs of the C, N, and O atoms which drive the gas-grain cycling process in our models (H$_2$O, N$_2$, CO, and CH$_4$) are shown in Fig. \ref{fig:H2O-GH2O-N2-GN2-CO-GCO}.

The ratio of the NH$_{3,{\rm{ice}}}$ to H$_2$O$_{\rm{ice}}$ abundances on the grains in our $\epsilon$ \text{=} 0.001 ALL model exceeds the upper limit reported by \citet{knez2005spitzer} starting at $\sim$3 $\times$ 10$^6$ years in the models which include CRDES. This is an effect of the highly efficient production of NH$_{3,{\rm{ice}}}$ from N atoms which have been derived from N$_{2,{\rm{ice}}}$ that has cycled back to the gas phase via cosmic-ray heating. Conversely, the contributions from H$_2$ desorption and cosmic-ray-induced photodesorption tend to reduce the quantity of NH$_{3,{\rm{ice}}}$ on the grains (with respect to that attained in the THERM model) between $\sim$4 $\times$ 10$^6$ and 3 $\times$ 10$^7$ years due to the fact that more NH$_{3,{\rm{ice}}}$ is directly re-injected into the gas phase via these mechanisms. The maximum difference in the NH$_{3,{\rm{ice}}}$ abundance calculated between the CRDES-dominated and H2DES models is $\sim$1.5 orders of magnitude.

The fact that the highest abundance of NH$_{3,{\rm{ice}}}$ occurs in the default ALL model (as well as in the CRDES model, Fig. \ref{fig:NH3-GNH3-epsilonVary}) is initially counterintuitive. With all of the desorption mechanisms in play, we might have expected the solid-state abundances of NH$_3$ to have become more or less reduced from the grains by the end of the chemical evolution. However, the reason that the NH$_{3, {\rm{ice}}}$ abundance prevails at late times in this model is because the efficient cycling of N$_{2, {\rm{ice}}}$ from the grains back to the gas phase through cosmic-ray heating induces the return of N atoms to the solid state through the aforementioned routes, thereby re-galvanising the production of NH$_{3, {\rm{ice}}}$ through the hydrogenation of N atoms. The fact that N$_{2, {\rm{ice}}}$ is generally one of the most abundant molecules on the grains in our models ensures that when its desorption is efficient, the transfer of N atoms from N$_2$ into NH$_3$ will be efficient as well, both in the gas phase and on the grains. Indeed, when cosmic-ray heating dominates, the primary late-time nitrogen reservoir switches from N$_{2, {\rm{ice}}}$ to NH$_{3, {\rm{ice}}}$. While H2DES and PDD tend to remove NH$_3$ from the grains, their effects on the NH$_{3,{\rm{ice}}}$ abundances are overpowered by CRDES for the parameters adopted here. Moreover, the re-injection of H$_2$O$_{\rm{ice}}$ into the gas phase through H2DES and PDD in fact bolsters the gas-phase production of NH$_3$ through the NO intermediate, thereby mitigating the loss of NH$_{3}$ from the system when H2DES and PDD are relevant. Network diagrams for the most important reactions involved in the formation of NH$_3$ in our models at 10$^{5}$ and 10$^{7}$ years are given in Figs. \ref{fig:NH3-formation-ALL-1e5yrs} and \ref{fig:NH3-formation-ALL-1e7yrs} in the Appendix.

\section{Summary and Conclusions}
\label{sec:Discussion}
In this work, we have presented and analysed the results of our newly-developed dark-cloud astrochemical models which utilise an updated version of the latest gas-phase reactions from the UDfA reaction network and include non-thermal desorption mechanisms as well as newly-implemented calculations for the formation of H$_2$ on a population of realistic interstellar grain types. We have compared our modelled results with observations towards TMC-1 (CP) and found that when employing the updated treatment for H$_2$ formation, the maximum agreements achieved by the C-class and N-class species are improved, whereas the maximum agreement achieved by the O-class species is suppressed. Regardless of the mechanism adopted for grain-surface H$_2$ formation, however, we find that the best agreements in our models are generally achieved when the species classes are considered separately rather than as a collective. The time dependency of the agreements for the different species classes in our models is consistent with the chemical effects that we might expect from the physical interactions of as-yet unresolved substructures within the cloud.





Due to its early-time synthesis from CN, we find that NH$_3$ can exist in the gas phase at its observed abundances at early times as well as late. Given the traditional interpretation of NH$_3$ as a "late-time" species, this carries implications for how we discern and implement observations of it in the context of the cloud's chemical evolution. Moreover, the variation in the gas-phase NH$_3$ abundance at late times in the ALL model suggests that the evaluation of species abundances relative to the NH$_3$ abundance might produce results which are more sensitive to the age of the source than previously considered, especially if the gas-grain cycling is dominated by cosmic-ray heating.



We focused our analysis of the chemistry on the cycling of molecular material between the gaseous and solid states at late times and found that while cosmic-ray heating tends to most effectively remove N$_2$, CO,  and other volatiles from the grains, H$_2$O is, depending on the efficiency chosen for the H2DES process, more directly affected by H$_2$ desorption. We saw that increasing the efficiency of H$_2$ desorption advances the abundance peak of gas-phase NH$_3$ to earlier times and reduces the abundance of solid-state NH$_{3,\rm{ice}}$ at late times. The finding that higher abundances of NH$_{3,\rm{ice}}$ are obtained for lower values of $\epsilon$ is the result of both the inefficient removal of the molecule from the grains through H$_2$ desorption when the $\epsilon$ value is low and the enhanced synthesis of the molecule in the solid state from N atoms derived from the gas-grain cycling of N$_2$ induced by cosmic-ray desorption. We found that for $\epsilon$ values of 0.05 or greater, the NH$_{3,\rm{ice}}$ abundance at late times is dominated by H$_2$ desorption, whereas for lower $\epsilon$ values, it is increasingly dominated by cosmic-ray desorption. Due to the dependency of H$_2$ desorption on the rate of H$_2$ formation as well, this result is inherently linked to the composition of the grains.

Given the unknowns inherent to the desorption mechanisms considered herein and the reality that we do not currently know for certain which of these mechanisms actually occur in star-forming environments (let alone their relative efficiencies), it is difficult to derive a quantitative conclusion about the chemistry based on our models. However, our results do indicate in a general sense that the synthesis of NH$_{3,\rm{ice}}$ -- and by extent, the synthesis of various complex organic ices -- is indeed sensitive to the nature of the gas-grain cycling of molecular material within the system. In the context of the chemical differentiation commonly observed in star-forming environments, our results suggest that the gas-phase synthesis of complex O-bearing organic molecules at later stages of the cloud's evolution should at least theoretically be limited when the cycling in the prestellar core is dominated by cosmic-ray heating (due to the enhanced quantities of NH$_{3,\rm{ice}}$ formed and stored on the grains) and enhanced when it is dominated by H$_2$ desorption (due to the reduced presence of NH$_{3,\rm{ice}}$ on the grains). However, this result also suggests that the solid-state production of nucleobases such as adenine and guanine could plausibly be enhanced by the increased abundances of NH$_{3}$ on the grains, at least when the cycling is dominated by cosmic-ray heating.


We thank the anonymous referee for their valuable comments and insights which improved the content of this manuscript. TJM is grateful to the STFC for support through grant ST/P000312/1.

\section{Data Availability}
\label{sec:data-availability}

The data underlying this article will be shared on reasonable request to the corresponding author.

\bibliographystyle{mnras}
\bibliography{SpeciesCycling2020-Revisions-Nchem3} 

\section{Appendix}
\label{sec:Appendix}
\begin{figure*}
   \resizebox{\hsize}{!}{\includegraphics{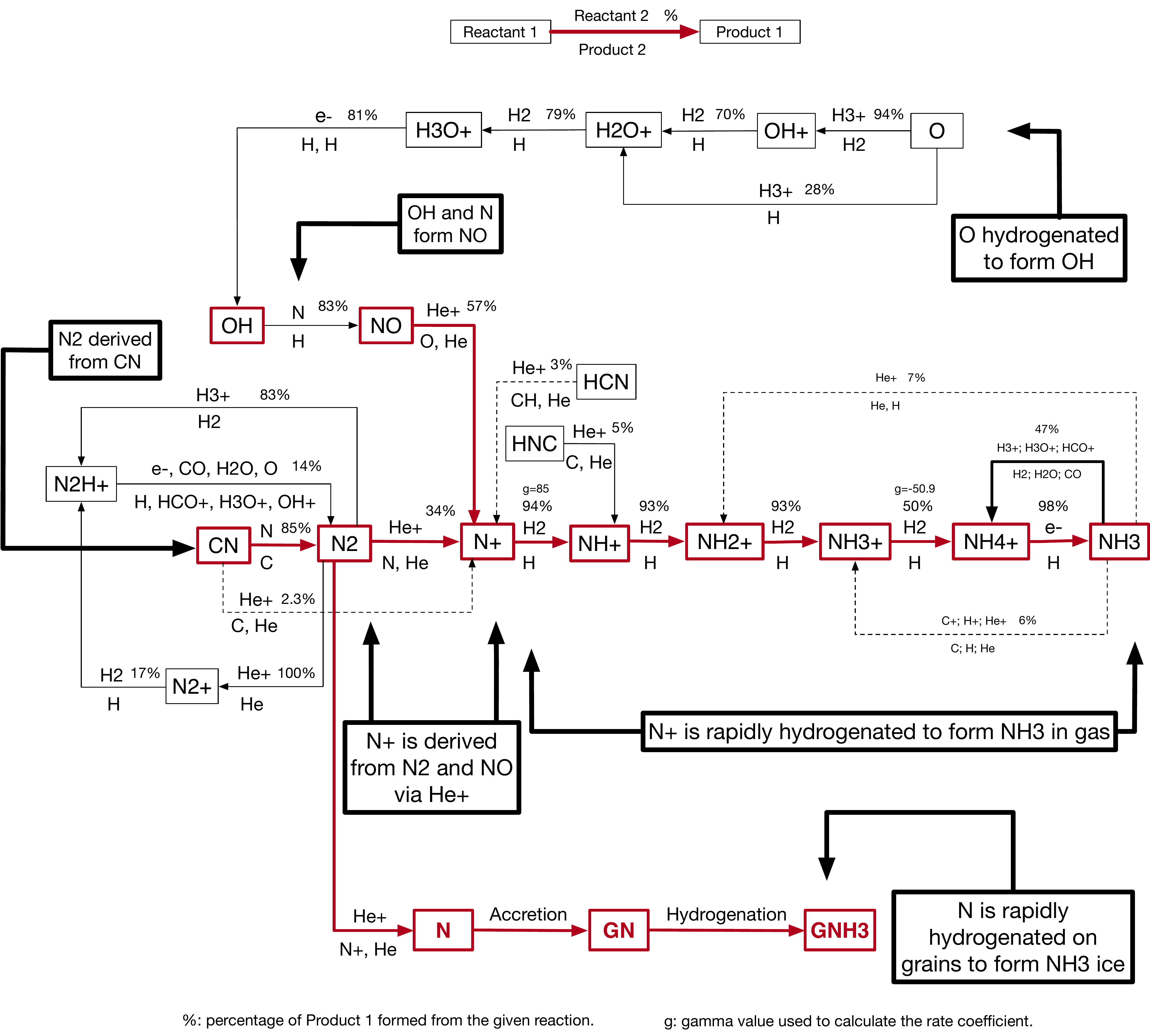}}   
  \caption[Model description as in Fig. \ref{fig:desorptions-NH3-HC3N}.]{\small{Formation of NH$_3$ in the gas and on the grains in our ALL ($\epsilon$ = 0.001) model at 10$^{5}$ years. }}
  \label{fig:NH3-formation-ALL-1e5yrs}
\end{figure*}

\begin{figure*}
   \resizebox{\hsize}{!}{\includegraphics{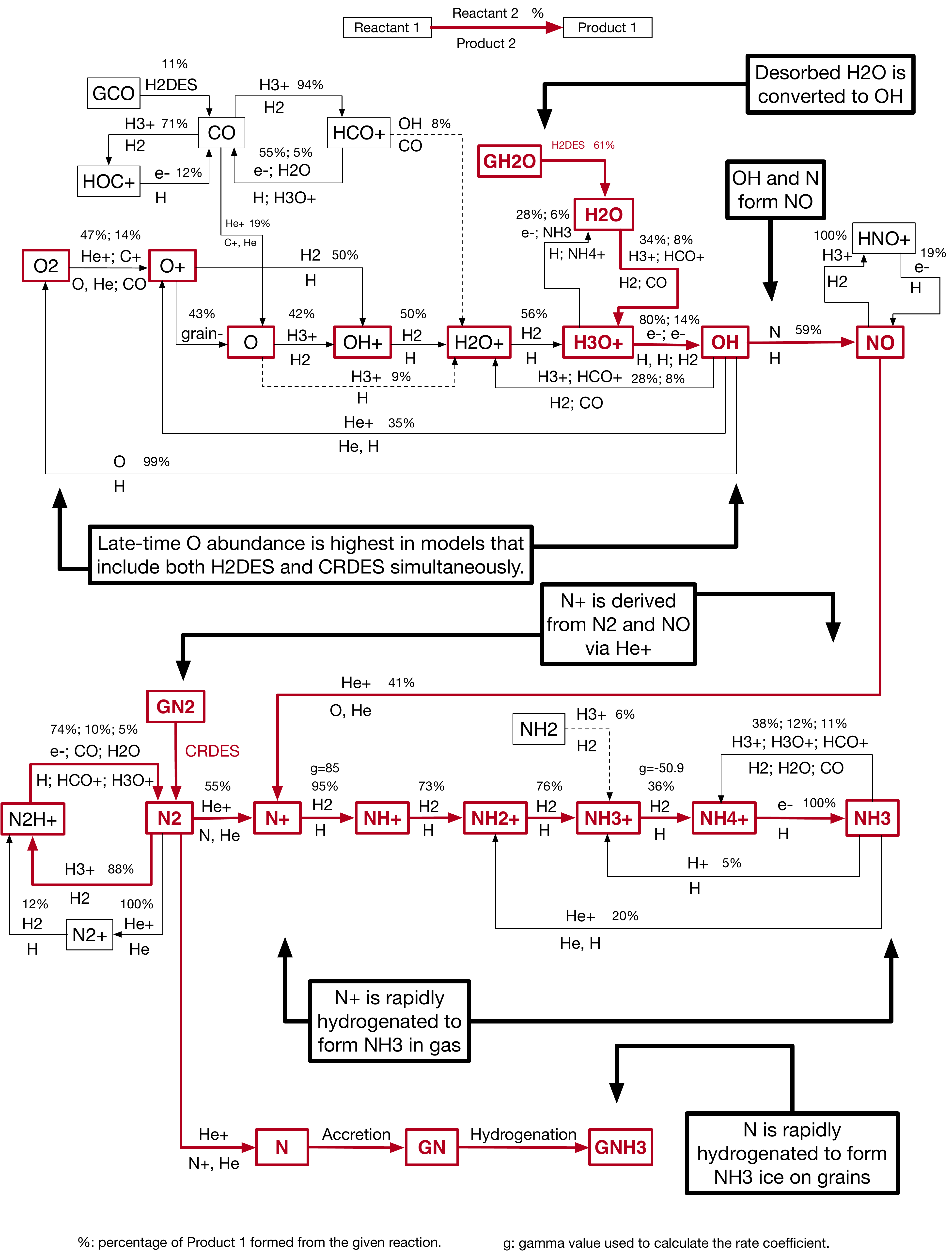}}   
  \caption[Model description as in Fig. \ref{fig:desorptions-NH3-HC3N}.]{\small{Formation of NH$_3$ in the gas and on the grains in our ALL ($\epsilon$ = 0.001) model at 10$^{7}$ years.}}
  \label{fig:NH3-formation-ALL-1e7yrs}
\end{figure*}

\bsp	
\label{lastpage}
\end{document}